\documentclass[twocolumn]{aastex62}

\usepackage{amsmath}
\usepackage{amssymb}
\usepackage{graphicx}
\usepackage{color}

\usepackage{savesym}
\savesymbol{tablenum}
\usepackage{siunitx}
\usepackage{xspace}
\restoresymbol{SIX}{tablenum}

\newcommand{\St}{\ensuremath{\mathrm{St}}\xspace}

\newcommand\dbquote[1]{\textquotedblleft #1\textquotedblright}
\newcommand\sgquote[1]{\textquoteleft #1\textquoteright}

\newcommand\partialdiff[1]{\frac{\partial}{\partial #1}}
\newcommand\stdiff[1]{\textrm{d} #1}

\graphicspath{{./}{figures/}}

\received{2018 October 12}
\revised{2018 November 23}
\accepted{2018 November 24}

\submitjournal{ApJ}

\shorttitle{Dust accretion preceding an outburst?}
\shortauthors{G\'arate et al.}

\begin{document}
\title{\Large{The dimming of RW Auriga. Is dust accretion preceding an outburst?}}

\correspondingauthor{Mat\'ias G\'arate}
\email{mgarate@usm.lmu.de}
\author[0000-0001-6802-834X]{Mat\'ias G\'arate}
\affiliation{University Observatory, Faculty of Physics, Ludwig-Maximilians-Universit\"at M\"unchen, Scheinerstr.\ 1, 81679 Munich, Germany}

\author[0000-0002-1899-8783]{Til Birnstiel}
\affiliation{University Observatory, Faculty of Physics, Ludwig-Maximilians-Universit\"at M\"unchen, Scheinerstr.\ 1, 81679 Munich, Germany}

\author[0000-0002-1589-1796]{Sebastian Markus Stammler}
\affiliation{University Observatory, Faculty of Physics, Ludwig-Maximilians-Universit\"at M\"unchen, Scheinerstr.\ 1, 81679 Munich, Germany}

\author[0000-0003-4243-2840]{Hans Moritz G\"unther}
\affiliation{MIT, Kavli Institute for Astrophysics and Space Research, 77 Massachusetts Avenue, Cambridge, MA 02139, USA}

\begin{abstract}
RW Aur A has experienced various dimming events in the last years, decreasing its brightness by $\sim 2\ \textrm{mag}$ for periods of months to years. Multiple observations indicate that a high concentration of dust grains, from the protoplanetary disk's inner regions, is blocking the starlight during these events. 
We propose a new mechanism that can send large amounts of dust close to the star on short timescales, through the reactivation of a dead zone in the protoplanetary disk.
Using numerical simulations we model the accretion of gas and dust, along with the growth and fragmentation of particles in this scenario.
We find that after the reactivation of the dead zone, the accumulated dust is rapidly accreted towards the star in around \SI{15}{years}, at rates of $\dot{M}_\textrm{d} = \SI{6e-6}{M_\odot/yr}$ and reaching dust-to-gas ratios of $\epsilon \approx 5$, preceding an increase in the gas accretion by a few years.
This sudden rise of dust accretion can provide the material required for the dimmings, although the question of how to put the dust into the line of sight remains open to speculation.
\end{abstract}


\section{Introduction} \label{sec_Intro}
RW Aur A is a young star that in the last decade presented unusual variations in its luminosity. The star has about a solar mass, it is part of a binary system, and is surrounded by a protoplanetary disk showing signatures of tidal interaction \citep{Cabrit2006,Rodriguez2018}.
The star had an almost constant luminosity for around a century, interrupted only by a few short and isolated dimmings (see \citet{Berdnikov2017} for a historical summary) until 2010, when its brightness suddenly dropped by \SI{2}{mag} in the V band for \SI{6}{months} \citep{Rodriguez2013}. 
Since 2010, a total of five dimming events have been recorded \citep[see][among others]{Rodriguez2013, Rodriguez2016, Petrov2015, Lamzin2017, Berdnikov2017}.
The dimmings can last from a few months to two years, and reduce the brightness of the star up to \SI{3}{mag} in the visual. Moreover, there is no obvious periodicity in their occurrence, and their origin is not yet clear \citep[a summary of the events can be found in ][]{Rodriguez2018}.
\subsection{Observations of RW Aur Dimmings} \label{sec_Intro_Observations}
Some observations in the recent years have shed light on the nature of RW Aur A dimmings:\\
During the event in 2014-2015 \citep{Petrov2015}, observations by \citet{Shenavrin2015} show a increase in IR luminosity at bands L and M. The authors infer that hot dust from the inner regions is emitting the infrared excess, while occluding the starlight and causing the dimming in the other bands.\\
Observations by \citet{Antipin2015,Schneider2015}
found that the absorption from optical to NIR wavelengths is gray, which could indicate the presence of large particles causing the dimming ($\gtrsim \SI{1}{\mu m}$), and measured a dust column density of \SI{2e-4}{g/cm^2}, although a similar absorption may be produced if an optically thick disk of gas and small grains partially covers the star \citep{Schneider2018}.\\
Then, the study of RW Aur A spectra by \citet{Facchini2016} also found that the inner accretion regions of the disk are being occluded, and therefore the dimmings should come from perturbations at small radii.\\
Finally, during the dimming in 2016, X-Ray observations from \citet{Gunther2018} indicate super-Solar Fe abundances, along with a higher column density of gas in the line of sight. The high $N_\textrm{H}/A_\textrm{V}$ found by the authors is interpreted as gas rich concentrations in the occluding material, or as a sign of dust growth. Also, gray absorption was found again during this event.\\
Given this information, several authors have discussed what mechanism would put the dust from the innermost regions of the protoplanetary disk, into the line of sight. Among the possible explanations are:
a warped inner disk \citep{Facchini2016,Bozhinova2016}, stellar winds carrying the dust \citep{Petrov2015,Shenavrin2015}, planetesimal collision \citep{Gunther2018}, and a puffed up inner disk rim \citep{Facchini2016,Gunther2018}.\\
Most of the proposed mechanisms rely on having enough dust close to the star to cause the dimmings. So the focus of this paper is to propose a mechanism that can deliver large amounts of dust to the inner regions of the protoplanetary disk, by raising the dust accretion rate through the release of a dust trap.\\
An increased dust concentration can explain some aspects of the dimmings, such as the high metallicity \citep{Gunther2018}, the emission of hot dust \citep{Shenavrin2015}, and cause the dimmings provided that another mechanism transports it into the line of sight.
\subsection{A Fast Mechanism for Dust Accretion}
A sudden rise in the dust accretion can occur in the early stages of stellar evolution, when the stars are still surrounded by their protoplanetary disk composed of gas and dust. 
The dynamics of the gas component of the disk is governed by the viscous evolution, which drives the accretion into the star \citep{Lynden-Bell1974}, and the pressure support, that produces the sub-keplerian motion. 
On the other hand, the dust particles are not affected by pressure forces, but suffer the drag force from the gas. This interaction extracts angular momentum from the particles and causes them to drift towards the pressure maximum, with the drift rate depending on the coupling between the dust and gas motion \citep{Whipple1972,Weidenschilling1977,Nakagawa1986}. 
This means that any bumps in the gas pressure act as concentration points for the dust. In these dust traps the grains accumulate, grow to larger sizes, and reach high dust-to-gas ratios \citep{Whipple1972, Pinilla2012}.\\
One of the proposed mechanisms to generate a pressure bump is through a dead zone \citep{Gammie1996}, a region with low turbulent viscosity (the main driver of accretion), due to a low ionization fraction which turns off the magneto-rotational instability \citep{Balbus1991}, that allows the gas to accumulate until a steady state is reached.\\
The presence of a dead zone would allow the dust to drift and accumulate at its inner border \citep{Kretke2009}, which can be located at the inner regions of the protoplanetary disk ($r\lesssim \SI{1}{au}$). Assuming that these conditions are met, the reactivation of the dead zone turbulent viscosity through thermal, gravitational or magnetic instabilities \citep{Martin2011} would break the steady state and allow all the accumulated material (both gas and dust), to be flushed towards the star. 
This mechanism has been invoked already in the context of FU Ori objects to explain the variability in their accretion rate through outbursts \citep{Audard2014}.\\
Since the dust is drifting faster than the gas, because of hydrodynamic drag and dust diffusion, it will arrive at the inner boundary of the disk first, where it may generate the observed hot dust signatures, Fe abundances, and the dimmings if it enters into the line of sight through either a puffed up inner rim, or a stellar wind (among the explanations mentioned above). Therefore, the accretion of large amounts of dust could be actually followed by an increase in the gas accretion rate.\\
In this study we use 1D simulations of gas and dust, including dust coagulation and fragmentation, to model the concentration of dust at the inner edge of a dead zone. Subsequent reactivation of the turbulent viscosity lets the accumulated material rapidly drift  towards the star. 
We measure the timescale required for the dust drifting, how much dust can be concentrated into the inner regions by this mechanism, and put it into the context of the observed features during the dimmings along with the possible explanations listed.\\
In \autoref{sec_Model} we define the relevant equations that dominate the gas and dust dynamics, and our model for the dead zone. 
In \autoref{sec_Setup} we describe the implementation of the accumulation phase and the reactivation phase of our simulations is discussed, and the relevant parameter space for the disk and dead zone properties are explored.
In \autoref{sec_Results} we show the results on the dust accretion rate towards the star and how the different properties of the disk and the consideration of dust back-reaction affect these results.
In \autoref{sec_Discussion} we compare our findings with the data from RW Aur and discuss how our model could be adjusted. Finally, we conclude in \autoref{sec_Summary}.
\newpage
\section{Model Description} \label{sec_Model}
We model a protoplanetary disk composed of gas and dust, solving the advection equation for both components. Our model includes multiple dust species and coagulation as in \citet{Birnstiel2010}, and the presence of a dead zone in the inner disk.
The advection of gas and dust is solved using the mass conservation equation:
\begin{equation} \label{eq_mass_conservation}
\partialdiff{t} \left(r \, \Sigma_{\textrm{g,d}}\right) + \partialdiff{r} (r \, \Sigma_{\textrm{g,d}} \, v_{\textrm{g,d}})= 0,
\end{equation}
where $r$ is the radial coordinate, $\Sigma$ is the mass surface density, $v$ is the radial velocity, and the subindex \sgquote{g} and \sgquote{d} refer to the gas and dust respectively.
\subsection{Viscous Evolution}
The gas is accreted towards the star following the viscous evolution theory, for which the viscous velocity is:
\begin{equation} \label{eq_visc_velocity}
v_\nu = -\frac{3}{\Sigma_\textrm{g} \sqrt{r}}\partialdiff{r}(\nu \, \Sigma_\textrm{g} \, \sqrt{r}),
\end{equation}
with $\nu$ the turbulent viscosity \citep{Lynden-Bell1974}. If we neglect the effect of dust back-reaction we can assume that the radial velocity of the gas is $v_\textrm{g,r} = v_\nu$.
We follow the \citet{Shakura1973} $\alpha$ model for the viscosity:
\begin{equation} \label{eq_alpha_visc}
\nu = \alpha \, c_s^{2} \, \Omega_K^{-1},
\end{equation}
where the $\alpha$ parameter controls the intensity of the viscosity (and therefore the accretion), $\Omega_K$ is the Keplerian angular velocity, and $c_s = \sqrt{\gamma K_b T/\mu m_\textrm{H}}$ is the adiabatic gas sound speed, with $T$ the gas temperature, $K_b$ the Boltzmann constant, $\mu = 2.3$ the mean molecular mass, $\gamma =1.4$ the adiabatic index, and $m_\textrm{H}$ the hydrogen mass.\\
As a starting point for our model, we assume that the gas is in steady state, such that the accretion rate $\dot{M}_\textrm{g}$ is constant over the disk radius following:
\begin{equation} \label{eq_steady_state}
3\,\pi \, \Sigma_\textrm{g} \, \nu = \dot{M}_\textrm{g}.
\end{equation}
\subsection{Gas Orbital Velocity}
The gas orbital motion is partly pressure supported, so for a negative pressure gradient it will orbit at sub-keplerian velocity. This point is particularly important for dust dynamics \citep{Whipple1972, Weidenschilling1977}.
The difference between the gas orbital velocity and the keplerian velocity, $\Delta v_{\textrm{g},\theta} = v_{\textrm{g},\theta} - v_K$, is:
\begin{equation} \label{eq_gas_orbitalVel}
\Delta v_{\textrm{g},\theta} = -\eta \, v_K,
\end{equation}
with $v_K$ the keplerian speed and
\begin{equation} \label{eq_eta}
\eta = -\frac{1}{2} \left(\rho_\textrm{g} \, r \, \Omega_K^2\right)^{-1} \frac{\partial P}{\partial r}. 
\end{equation}
Here $\rho_\textrm{g}$ is the gas volume density in the disk's midplane, and $P=\rho_\textrm{g}c_s^2/\gamma$ is the pressure.
In \autoref{sec:backreaction}, we will consider the effects of the dust back-reaction on the gas orbital velocity.
\subsection{Dead Zone Model}
\begin{figure}
\centering
\includegraphics[width=80mm]{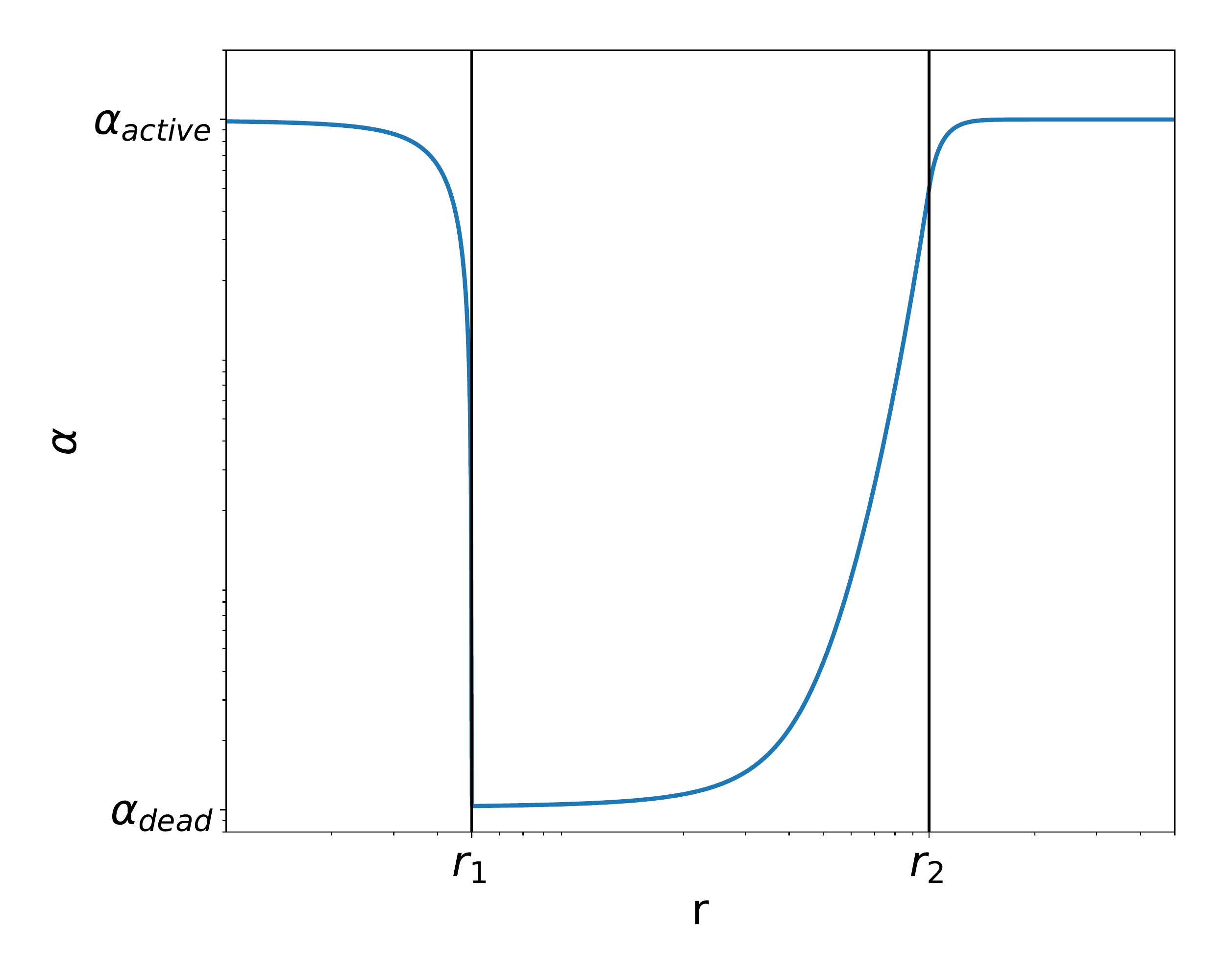}
 \caption{The diagram shows the shape of the $\alpha$ radial profile for our dead zone model (in logarithmic scale). This consists on three regions, the active inner zone limited by a sharp decay at $r_1$, the dead zone with a smooth rise towards its outer edge around $r_2$, and the outer active zone extending until the outer boundary of the simulation.}
 \label{Fig_AlphaDiagram}
\end{figure}

Magneto-hydrodynamical models have predicted the presence of a region with low turbulence at the inner regions of protoplanetary disks, commonly called \dbquote{Dead Zone}, caused when the ionization fraction is too low for the magneto-rotational instability (MRI) to operate \citep{Gammie1996}.\\
We parametrize the dead zone by using a variable $\alpha$ parameter over radius, while remaining agnostic to the underlying physics, similar to previous research \citep{Kretke2009,Pinilla2016}. Our profile is defined as:
\begin{equation}\label{eq_alpha_profile}
\alpha(r) = \begin{cases}
	       \alpha_\textrm{active} - \Delta \alpha \cdot e^{5(\frac{r}{r_1}-1)}		&	r < r_1 \\       
           \alpha_\textrm{dead} + \Delta \alpha \cdot \frac{1}{2} e^{10(\frac{r}{r_2}-1)}    &    r_1 < r < r_2\\
           \alpha_\textrm{dead} + \Delta \alpha \cdot (1 - \frac{1}{2} e^{-10(\frac{r}{r_2}-1)})   &     r_2 < r,
      \end{cases}
\end{equation}
where $\alpha_\textrm{dead}$ and $\alpha_\textrm{active}$ are the characteristic values of the $\alpha$ parameter inside and outside the dead zone, $r_1$ and $r_2$ are its inner and outer edges, and $\Delta \alpha = \alpha_\textrm{active} - \alpha_\textrm{dead}$.
Although the dead zone shape is rather arbitrary, it was chosen to have a smoother outer border than the one of \citep{Kretke2009}, but retaining a sharp inner edge where the dust accumulates.
\autoref{Fig_AlphaDiagram} shows a diagram of the $\alpha$ profile, illustrating the shape of the different intervals and its main components to guide the reader with \autoref{eq_alpha_profile}.\\
In order to maintain the steady state from \autoref{eq_steady_state} we define the initial surface density and temperature profiles as follow:
\begin{equation} \label{eq_gas_ini_dens}
\Sigma_\textrm{g}(r) = \Sigma_\textrm{0} \left( \frac{r}{r_0} \right)^{-1} \frac{\alpha_\textrm{active}}{\alpha(r)},
\end{equation}
\begin{equation} \label{eq_gas_ini_T}
T(r) = T_0 \left( \frac{r}{r_0} \right)^{-1/2},
\end{equation}
where $\Sigma_0$ and $T_0$ are the values of the density and temperature at $r_0 = \SI{1}{au}$.
\subsection{Dead Zone Reactivation}
While the dead zone is present the gas will remain in steady state and the dust will accumulate at its inner boundary. 
Yet, different processes can reactivate the turbulence in the dead zone, allowing the accumulated material to flush towards the star. In the Gravo-Magneto instability \citep{Martin2011} for example, the gas in the dead zone becomes gravitationally unstable, raising the temperature to the point of triggering the MRI, and finally producing an accretion outburst.\\
In our simulations we remain agnostic about the mechanism that causes the reactivation, and only set the reactivation time $t_\textrm{r}$ arbitrarily, such that:
\begin{equation} \label{eq_alpha_reactive}
\alpha(r,\, t> t_\textrm{r}) = \alpha_\textrm{active}. 
\end{equation}

\subsection{Dust Dynamics}
The dust dynamics are governed by the gas motion and the particle size. The time required for a particle of size $a$ and material density $\rho_s$ to couple to the motion of the gas is called the stopping time, and is defined as:
\begin{equation} \label{eq_tstop}
t_\textrm{stop} = \begin{cases}
					\sqrt{\frac{\pi}{8}}\frac{\rho_s}{\rho_\textrm{g}}\frac{a}{c_s} & \lambda_\textrm{mfp}/a \geq 4/9\\
					\frac{2}{9}\frac{\rho_s}{\rho_\textrm{g}}\frac{a^2}{\nu_\textrm{mol}} & \lambda_\textrm{mfp}/a < 4/9,\\
      				\end{cases}
\end{equation}
with the mean free path $\lambda_\textrm{mfp} = (n \sigma_{\textrm{H}_2})^{-1}$ (where $n$ is the number density, and $\sigma_{\textrm{H}_2} = \SI{2e-15}{cm^2}$), and the molecular viscosity $\nu_\textrm{mol} = \sqrt{2/\pi} c_s \lambda_\textrm{mfp}$ \citep[following the definitions in][]{Birnstiel2010}.\\
A more useful quantity to describe the dust-gas coupling is the Stokes number (or dimensionless stopping time), which is defined as:
\begin{equation} \label{eq_Stokes}
\mathrm{St}  = t_\textrm{stop}\Omega_K.
\end{equation}
From this quantity we can quickly infer if a dust grain is coupled ($\St \ll 1$) or decoupled ($\St \gg 1$) to the gas. For the midplane this can be rewritten as:
\begin{equation} \label{eq_Stokes_Mid}
\mathrm{St} = \begin{cases}
				\frac{\pi}{2} \frac{a \rho_s}{\Sigma_\textrm{g}} & \lambda_\textrm{mfp}/a \geq 4/9\\
                \frac{2 \pi}{9} \frac{a^2 \rho_s}{\lambda_\textrm{mfp} \Sigma_\textrm{g}} & \lambda_\textrm{mfp}/a < 4/9.
				\end{cases} 
\end{equation}
The dust radial velocity is given in \citet{Nakagawa1986,Takeuchi2002} as:
\begin{equation} \label{eq_dust_velocity}
v_\textrm{d} =  \frac{1}{1+St^2} v_{\textrm{g},r} + \frac{2 St}{1+St^2}\Delta v_{\textrm{g},\theta} - D_\textrm{d} \frac{\Sigma_\textrm{g}}{\Sigma_\textrm{d}} \partialdiff{r}(\frac{\Sigma_\textrm{d}}{\Sigma_\textrm{g}}).
\end{equation}
Here, the first term of the dust velocity is responsible for small grains to move along with the gas, while the second term is responsible for the dust to drift towards the pressure maximum. The last term corresponds to the dust diffusion contribution \citep[see][]{Birnstiel2010}, with $D_\textrm{d}$ the dust diffusivity defined following \citet{Youdin2007} as:
\begin{equation} \label{eq_dust_diffusion}
D_\textrm{d} = \frac{\nu}{(1+ \mathrm{St}^2)}.
\end{equation}
The dust coagulation model is specified in \citet{Birnstiel2010}. Since our model focus on the inner regions of the protoplanetary disks, the grain growth will always be limited by the fragmentation barrier \citep{Brauer2008}, and the maximum grain size will be approximately:
\begin{equation} \label{eq_frag_limit}
St_{\textrm{frag}} = \frac{1}{3}\frac{v_\textrm{frag}^2}{\alpha c_s^2},
\end{equation}
where $v_\textrm{frag}$ is the fragmentation velocity for dust particles \citep{Birnstiel2012}. For silicates this corresponds to $v_\textrm{frag} \approx \SI{1}{m/s}$ \citep{Guttler2010}.\\

\subsection{Dust Back-reaction Effects}
In protoplanetary disks, where the dust-to-gas ratio $\epsilon = \rho_\textrm{d}/\rho_\textrm{g}$ is assumed to be $0.01$, the effect of the dust onto the gas is often neglected. However, in regions with higher concentrations of solids, like in pressure bumps \citep{Pinilla2012}, the angular momentum transfered from the dust into the gas might be significant enough to alter its dynamics.\\
Many authors have already derived and included the effect of back-reaction (coupled with viscous evolution) into numerical simulations, showing in which regimes it should be considered \citep{Tanaka2005,Garaud2007,Kretke2009,Kanagawa2017,Taki2016,Onishi2017,Dipierro2018}.\\
In this paper we include the dust back-reaction in one of our simulations to study its impact on our model. To do so, the gas velocities $v_{\textrm{g},r}$ and $\Delta v_{\textrm{g},\theta}$ are rewritten as follows:
\begin{equation} \label{eq_backreaction_vr}
v_{\textrm{g},r} = A v_\nu + 2 B \eta v_K,
\end{equation}
\begin{equation} \label{eq_backreaction_vtheta}
\Delta v_{\textrm{g},\theta} = \frac{1}{2} B v_\nu - A \eta v_K.
\end{equation}
The back-reaction coefficients $A$ and $B$ measure the degree to which the gas dynamics are affected by the dust. These depend on the size distribution of the solids, and the dust-to-gas ratio.\\
The formal definition and origin of back-reaction coefficients can be found in the \autoref{sec_Appendix_BackReaction}, along with a quick interpretation of them.\\
At this point we only want to remark that in the limit where $\epsilon = 0$ we obtain $A = 1$ and $B = 0$, recovering the traditional velocities for the gas. When dust is present, the value of $A$ decreases and $B$ increases. 
Thus, from \autoref{eq_backreaction_vr} we see that back-reaction slows down the viscous evolution with the term $A v_\nu$, and pushes the gas outward with the term $2 B \eta v_K$.\\
We find this contracted notation specially useful to summarize the back-reaction contribution to the gas dynamic.
%

\section{Simulation Setup} \label{sec_Setup}
%
In this section we describe the observational constrains relevant for RW Aur A, the free parameters of our model, and the setup of our 1D simulations using the \texttt{twopoppy} \citep{Birnstiel2012} and \texttt{DustPy}\footnote{DustPy is a new Python code that solves the diffusion-advection of gas and dust, and the coagulation-fragmentation of dust, based on the \citet{Birnstiel2010} algorithm.} (Stammler $\&$ Birnstiel, in prep.) codes.\\
Our setup consists of three phases, the first phase simulates the dust accumulation at the dead zone, using a global disk simulation over long timescales ($\sim \SI{e5}{yr}$), but with a simplified and fast computational model for the dust distribution using only two representative populations. 
As the first phase only tracks the evolution of the surface density, in a second phase we recover the quasi-stationary particle size distribution at the inner disk ($r\leq \SI{5}{au}$) by simulating the dust growth and fragmentation of multiple dust species. 
Finally, the third phase simulates evolution of gas and dust (including coagulation, fragmentation, and transport) in the inner disk after the dead zone is reactivated, to study the accretion of the accumulated material towards the star over short timescales, and delivering the final results.\\
This setup is useful to save computational time, as we are interested only in the inner disk after the dead zone reactivation, but require the conditions given by the global simulation.

\subsection{Observational Constrains} \label{sec_Setup_Constrains}
RW Aur A is a young star with a stellar mass of $M_* = \SI{1.4}{M_\odot}$ \citep{Ghez1997,Woitas2001}. The circumstellar disk has an estimated mass around $M_\textrm{disk} \approx \SI{4e-3}{M_\odot}$ \citep{Andrews2005}, presents a high accretion rate of $\dot{M} \approx \SI{4e-8} - \SI{2e-7}{M_\odot/yr}$ \citep{Hartigan1995, Ingleby2013,Facchini2016}, and extends from a distance of $\sim \SI{0.1}{au}$ \citep{Akeson2005, Eisner2007} until \SI{58}{au} \citep{Rodriguez2018}.\\
For the temperature profile we use $T_0 = \SI{250}{K}$, which gives similar values to the \citet{Osterloh1995} profile in the inner regions of the disk for our choice of slope.\\
Using these parameters, \autoref{eq_alpha_visc} and \autoref{eq_steady_state}, we can constraint the values for the density and turbulence.
From the disk accretion rate, mass and size we infer the value for the density $\Sigma_0 = \SI{50}{g/cm^2}$ at $r_0 = \SI{1}{au}$, and the viscous turbulence $\alpha_\textrm{active} = 0.1$. These parameters yield values of $\dot{M}_\textrm{g} = \SI{5e-8}{M_\odot/yr}$ for the accretion rate, and $M_\textrm{disk} = \SI{2e-3}{M_\odot}$ for the disk mass (without considering the accumulation excess in the dead zone).
The turbulence parameter $\alpha_\textrm{active}$ used in our simulations is high, but necessary in order to account for the high accretion rates measured. 
%
\subsection{Phase 1: Dust Concentration at the Dead Zone} \label{sec_Setup_Phase1}
\begin{figure}
\centering
\includegraphics[width=80mm]{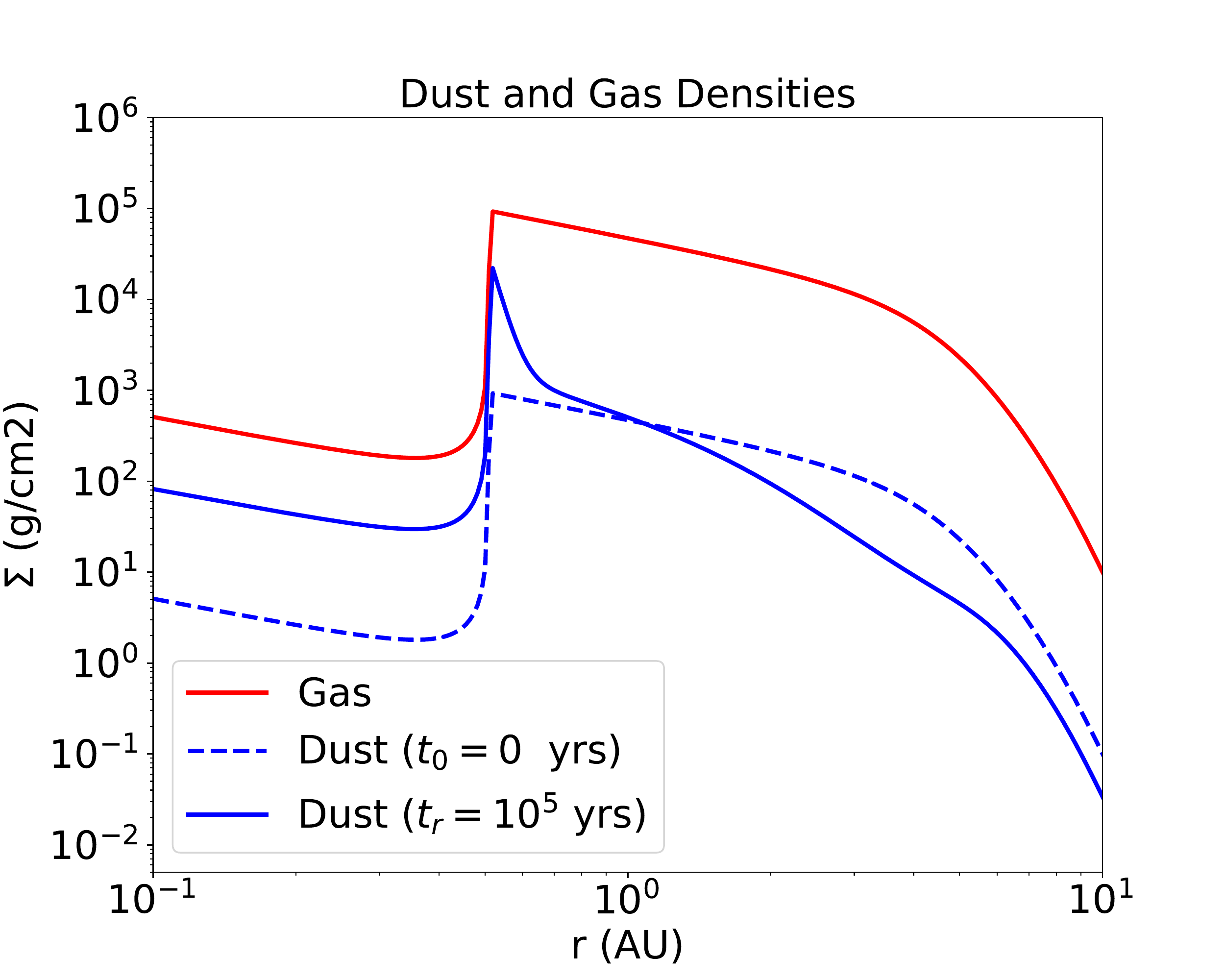}
 \caption{Gas and dust surface density obtained from TwoPopPy, at the beginning and at the end of the dust concentration phase. The gas (red) remains in steady state during this phase. The dust is initialized with a dust-to gas ratio of $\epsilon = 0.01$ (dashed blue line). The dust component is evolved for $\SI{e5}{yrs}$ (solid blue line) in which the dust concentrates at the inner disk, reaching $\epsilon \approx 0.24$ at the inner boundary of the dead zone, and $\epsilon \approx 0.16$ inside this region $r< r_1 = \SI{0.51}{au}$.
 }
 \label{Fig_TwoPoppyResult}
\end{figure}

In the first phase of our simulations we model the accumulation of dust in a disk with a dead zone, to obtain the dust-to-gas ratio radial profile.\\
We use the TwoPopPy code to simulate a global protoplanetary disk with two representative populations of the dust species \citep[details of the model can be found in][]{Birnstiel2012}. 
We initialize our simulations using \autoref{eq_alpha_profile}, \autoref{eq_gas_ini_dens} and \autoref{eq_gas_ini_T} for the $\alpha$ parameter, surface gas density and temperature profiles, with the values provided by the observational constrains. For the dust-to-gas ratio we assume an uniform initial value of $\epsilon = 0.01$.\\
For this phase, the simulation domain goes from $r_\textrm{in} = \SI{0.01}{au}$ to $r_\textrm{out} = \SI{100}{au}$, using $n_r = 500$ radial grid cells with logarithmic spacing. In the fiducial model, the inner and outer boundaries of the dead zone are $r_1 = \SI{0.51}{au}$, $r_2 = \SI{10}{au}$, with a depth of $\alpha_\textrm{dead} = \SI{e-4}{}$.\\
The simulation is evolved with this setup until the reactivation time $t_\textrm{r} = 10^5 \textrm{yrs}$. Approximately at this point the dust reaches its maximum accumulation at the inner boundary of the dead zone, which will yield the maximum dust accretion rate in the next phase. Since the gas is in steady state, we only evolve the dust in order to minimize possible numerical errors. Inside the dead zone, the gas phase is (marginally) gravitationally stable, with a Toomre parameter $Q = c_s \Omega_K/(\pi G \Sigma_\textrm{g}) \gtrsim 1.5$ \citep{Toomre1964}. A low Q value in this region does not conflict with the model, since the gravitational instability is one of the mechanisms that can eventually reactivate the dead zone.\\
The initial and final states of phase 1 are shown in \autoref{Fig_TwoPoppyResult}. During this phase the dust drifts towards the dead zone inner edge reaching values of $\epsilon = 0.24$, and concentrating \SI{110}{M_\oplus} between \SI{0.51}{} - \SI{0.6}{au}. 
Due to diffusion, the dust concentration at the innermost part of the disk also increases to values up to $\epsilon \approx 0.16$.
\subsection{Phase 2: Dust Size Distribution at the Inner Disk}  \label{sec_Setup_Phase2}
\begin{figure}
\centering
\includegraphics[width=80mm]{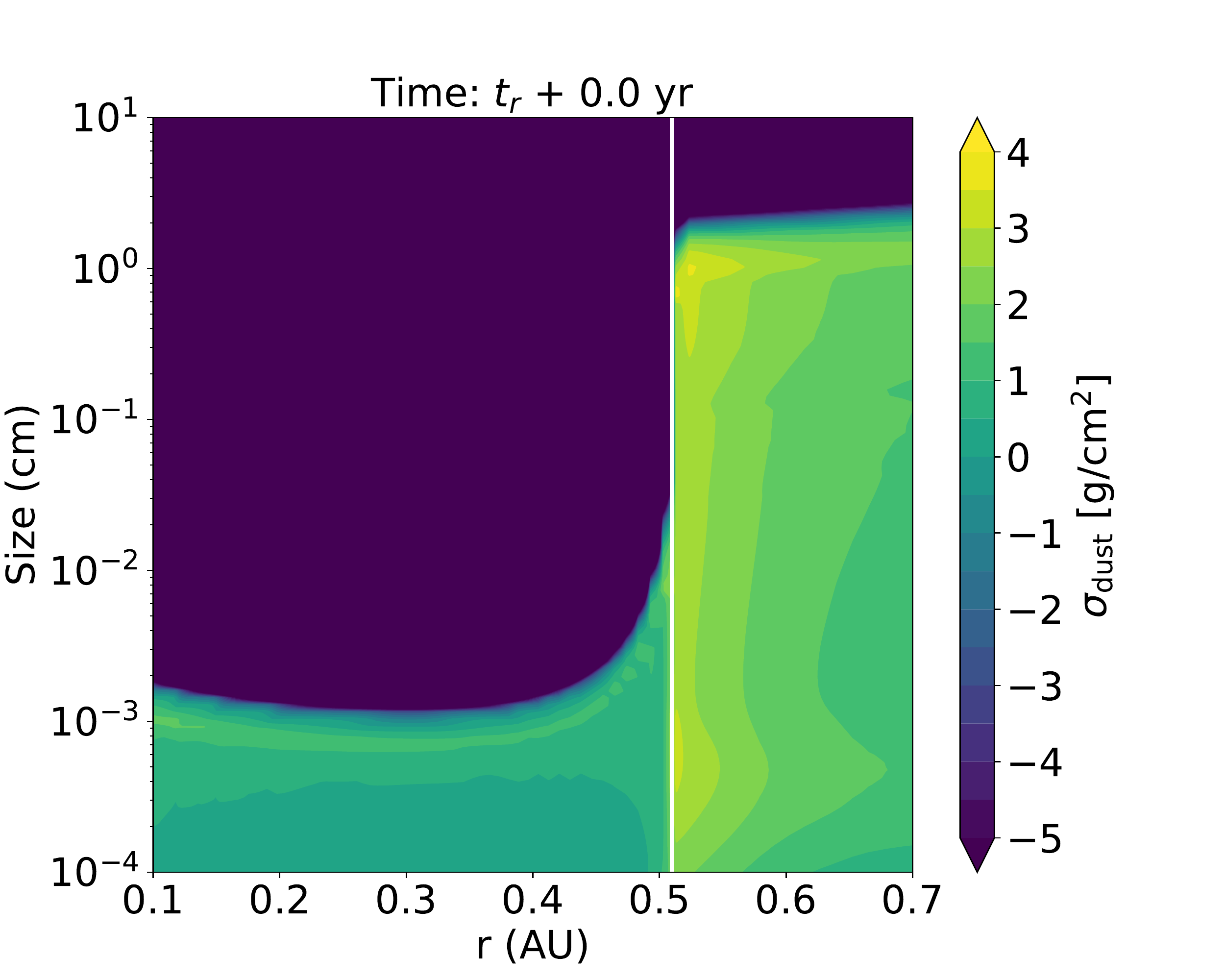}
 \caption{Dust distribution in the inner region of the protoplanetary disk immediately before the dead zone reactivation ($t = t_\textrm{r}$). In the dead zone, where the turbulence and the collision speed of solids are lower, the dust particles can grow to larger sizes  ($a_\textrm{max} \sim \SI{1}{cm}$) before reaching the fragmentation limit. At the active zone, the particles are respectively smaller ($a_\textrm{max} \sim \SI{10}{\mu m}$). The inner edge of the dead zone (marked by the white line) presents a high concentration of large dust grains.
 }
 \label{Fig_DustDistribution_Initial}
\end{figure}

In the second phase we want to recover the dust size distribution for multiple species, based on the dust-to-gas ratio and disk conditions obtained in the previous section.\\
We take the outcome of the TwoPopPy simulation as the new initial conditions, and use the DustPy code to solve the dust coagulation and fragmentation \citep[following the study of ][]{Birnstiel2010} at the inner disk, while \dbquote{freezing} the simulation exactly at the reactivation time $ t= t_r$, while the dead zone is still present (i.e. still using \autoref{eq_alpha_profile}). \\
The mass grid consists of $n_m=141$ logarithmic-spaced cells, between $m = \SI{e-15}{} - \SI{e5}{g}$, at every radius. 
Since in this phase we only care about the inner disk, we adjust our simulation radial domain to be from $r_\textrm{in} = \SI{0.05}{au}$ to $r_\textrm{out} = \SI{5}{au}$. The radial grid is defined as follow:
\begin{itemize}
\item 25 linear-spaced grid cells at $r = \SI{0.05}{} - \SI{0.09}{au}$,
\item 120 logarithmic-spaced grid cells at $r = \SI{0.09}{} - \SI{1.0}{au}$,
\item 20 logarithmic-spaced grid cells at $r = \SI{1.0}{} - \SI{5.0}{au}$.
\end{itemize}
The innermost region is necessary to avoid numerical problems with the inner boundary conditions. For optimization purposes we also turn off coagulation for $r < \SI{0.09}{au}$, since the growth and fragmentation timescales are so short in this region that the simulation becomes computationally unfeasible. Moreover, according to \citet{Akeson2005,Eisner2007} the inner boundary of RW Aur A disk should be around $r\sim \SI{0.1}{} - \SI{0.2}{au}$.
For these reasons all our analysis will only focus on the region of interest between $r = \SI{0.1}{} - \SI{1.0}{au}$.\\
We interpolate the gas and dust surface densities from the TwoPopPy simulation into the new grid, and use the coagulation model of DustPy to obtain the corresponding size distribution of the particles at $t = t_\textrm{r}$ for every radius. The dust distribution obtained at this phase is shown in \autoref{Fig_DustDistribution_Initial}, where the grains adjust to the fragmentation limit in the dead and active zones.

\subsection{Phase 3: Dead Zone Reactivation}  \label{sec_Setup_Phase3}
For the final phase we simulate the evolution of dust and gas in the inner disk, after the reactivation of the dead zone ($t > t_\textrm{r}$).\\
Once again we use the DustPy code, this time to solve the advection of gas and dust, along with the dust coagulation-fragmentation. We start this phase from the conditions given at Section \ref{sec_Setup_Phase2}, using the same grid for mass and radius, but now with the reactivated turbulence following \autoref{eq_alpha_reactive}. We let the simulation evolve for $\SI{15}{yrs}$, in which we expect that the material accumulated at the inner boundary of the dead zone will drift towards the star. 
The results of this phase on the accretion rate of gas and dust, as well as the final dust distribution,  will be shown in \autoref{sec_Results}.\footnote{The simulation data files and a plotting script are available in zenodo: \href{https://doi.org/10.5281/zenodo.1495061}{doi.org/10.5281/zenodo.1495061}.}

\subsection{Parameter Space}  \label{sec_Setup_ParSpace}
In \autoref{TableParam_Fiducial} we summarize the parameters used for the disk setup of our fiducial simulation.
As the properties of the dead zone are free parameters, chosen to be in a relevant range for the RW Aur dimming problem, we also require to explore (even briefly) the parameter space for these properties, and see how they affect the final outcome of the simulations. We present five additional simulations, changing one parameter of the fiducial model at a time, this way we explore the effect of having: no initial dust accumulation at reactivation, different dead zone properties, and the expected effects of back-reaction in the final result. The parameter changes are described in \autoref{TableParam_Variation}.

\begin{table}
\begin{center}
 \caption{Fiducial simulation parameters.}
 \label{TableParam_Fiducial}
  \begin{tabular}{ l  l }
    \hline
    \hline
    Parameter & Value  \\
    \hline
    $\Sigma_0$ & \SI{50}{g/cm^2}  \\
    $T_0$ & \SI{250}{K}  \\
    $r_0$ & \SI{1}{au}  \\
    $\alpha_\textrm{active}$ & \SI{e-1}{}  \\
    $\alpha_\textrm{dead}$ & \SI{e-4}{}  \\
    $r_1$ & \SI{0.51}{au}  \\ 
    $r_2$ & \SI{10}{au}  \\
    $t_\textrm{r}$ & \SI{e5}{yrs}  \\
    Back-reaction & Off  \\
    \hline
  \end{tabular}
  \end{center}
\end{table}

\begin{table}
\begin{center}
 \caption{Parameter variations.}
 \label{TableParam_Variation}
  \begin{tabular}{ l  l }
    \hline
    \hline
    Simulation & Parameter Changed  \\
    \hline
    Control Simulation & $t_\textrm{r} = \SI{0}{yrs}$  \\
    Shallow Dead zone  &  $\alpha_\textrm{dead} = \SI{e-3}{}$  \\
    Closer Inner Edge & $r_1 = \SI{0.25}{au}$  \\ 
    Closer Outer Edge & $r_2 = \SI{4}{au}$ \\
    Back-reaction & On  \\
    \hline
  \end{tabular}
  \end{center}
\end{table}

\section{Results} \label{sec_Results}
%

\begin{figure}
\centering
\includegraphics[width=80mm]{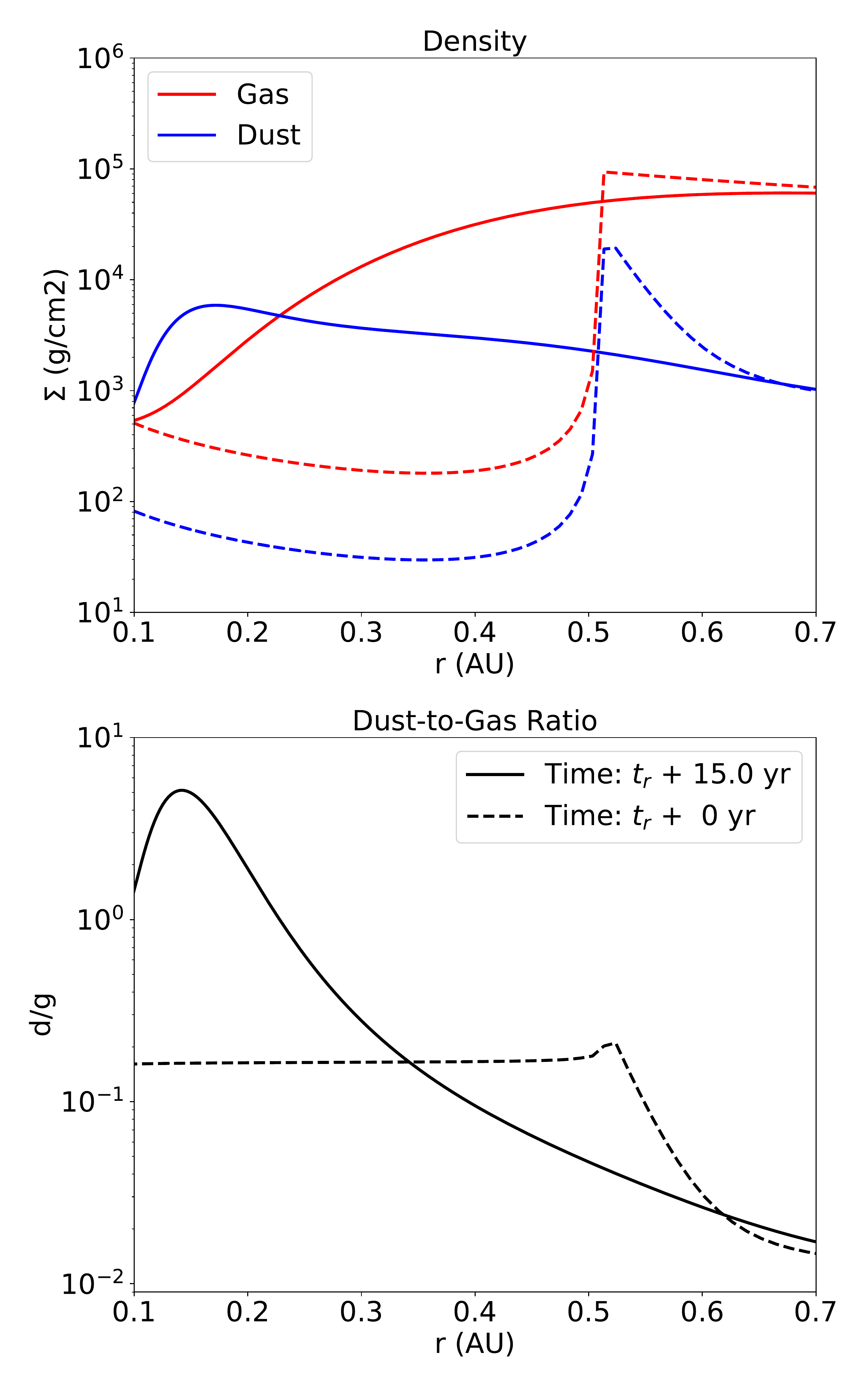}
 \caption{The plots show the simulation state immediately before the dead zone reactivation at $t=t_\textrm{r}$ (dashed lines), and \SI{15}{yrs} after it (solid lines).
\textit{Top}: Evolution of the gas (red) and dust (blue) surface densities. The initial state shows the gas steady state profile and the accumulation of dust at the inner boundary of the dead zone. After reactivation the accumulation of dust flushes towards the star faster than the gas.
\textit{Bottom}: Dust-to-gas ratio evolution. At the initial state the inner region presents an already high solid concentration thanks to mixing at the dead zone boundary. During the flushing the dust-to-gas ratio reaches values of $\epsilon = 5$ at some of the regions where the dust concentration arrived before the gas.
 }
 \label{Fig_FiducialFlushing}
\end{figure}

\begin{figure}
\centering
\includegraphics[width=80mm]{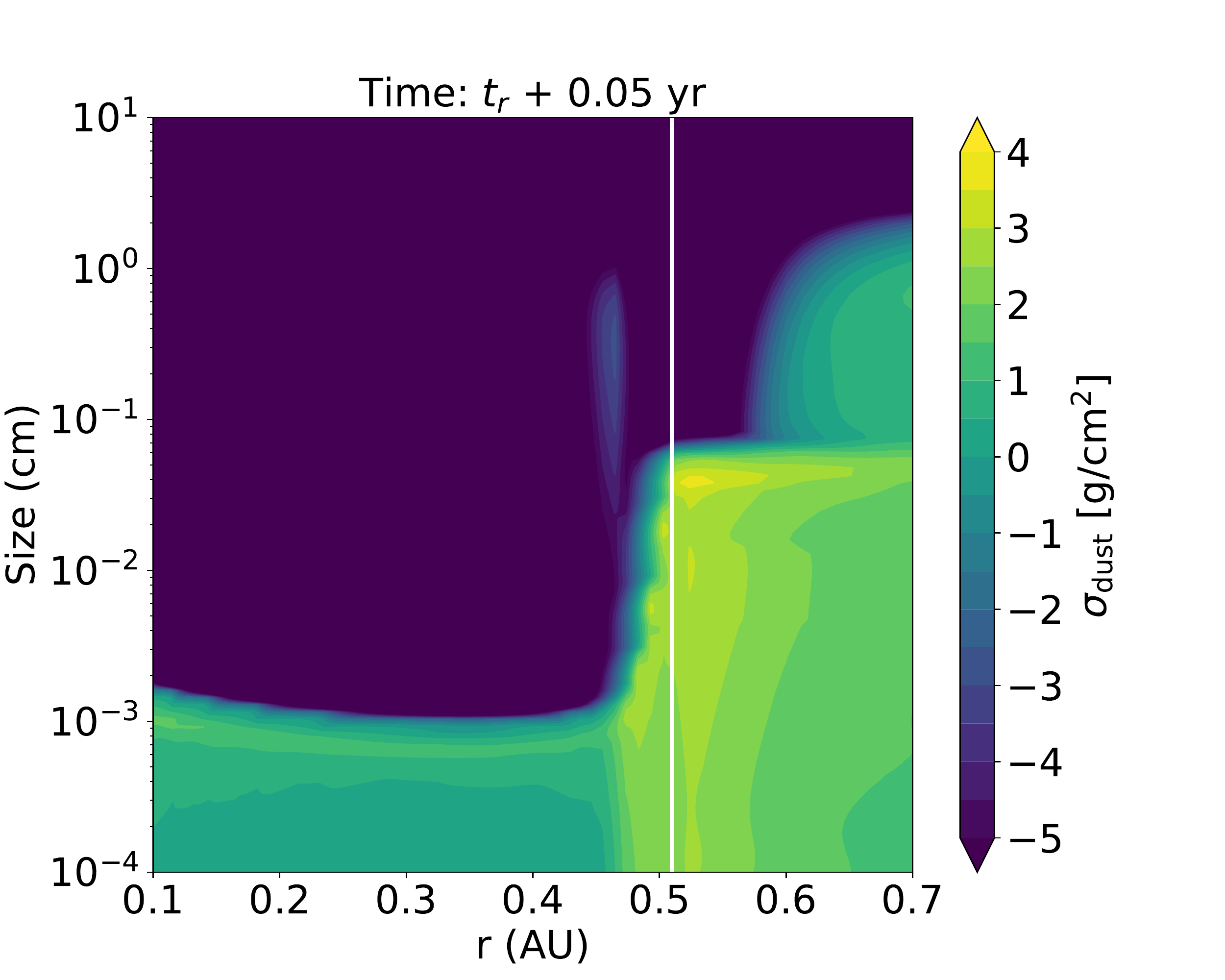}
\includegraphics[width=80mm]{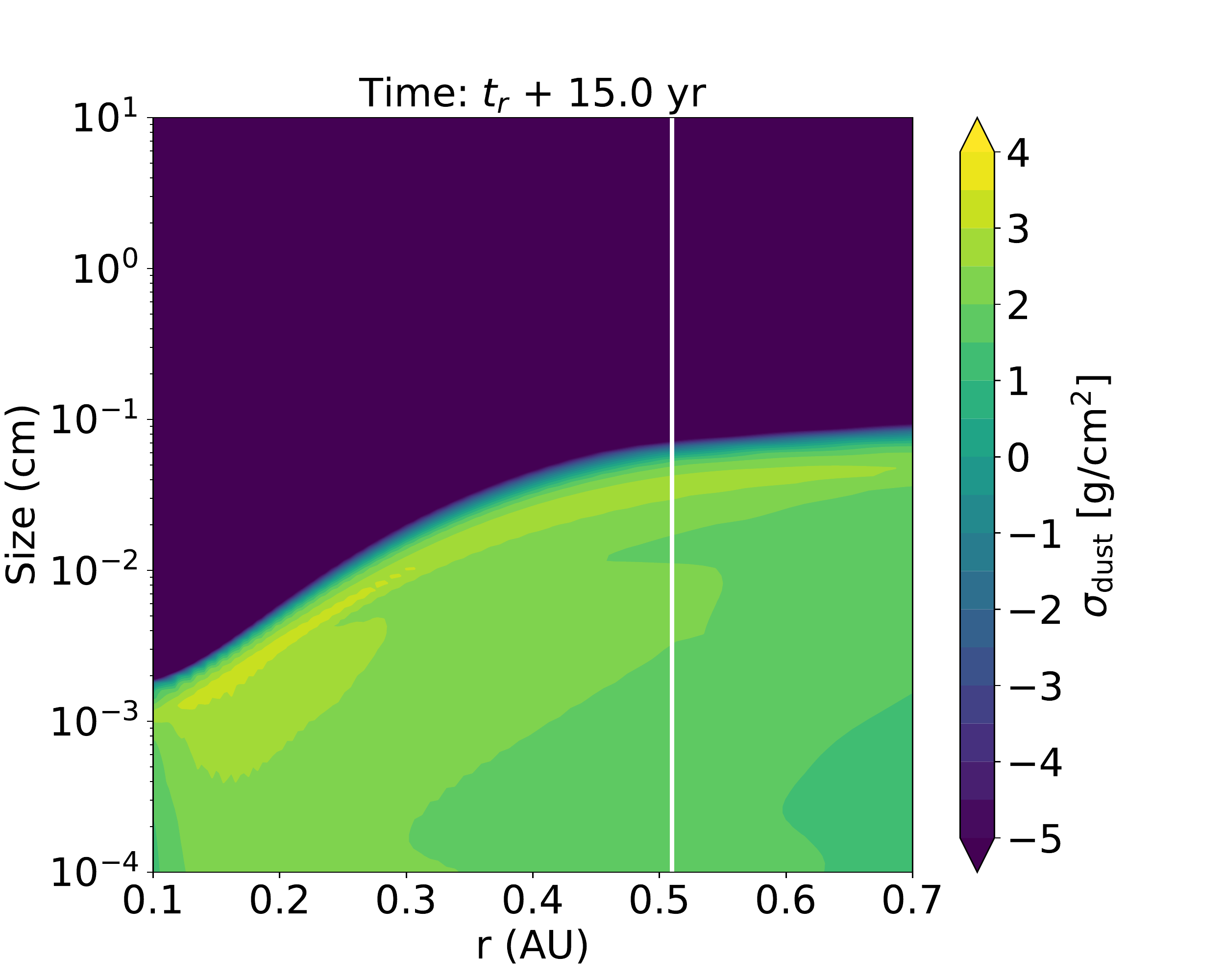}
 \caption{Dust distribution in the inner region of the protoplanetary disk after $\SI{0.05}{} \textrm{ and } \SI{15}{yrs}$ of the dead zone reactivation. \textit{Top}: The dust that was accumulated at the dead zone diffuses to the inner region within $\sim \SI{10}{}$ collisional times, generating high dust-to-gas concentrations. The original edge of the dead zone is marked in white. \textit{Bottom}: Afterwards, the dust drifts towards the inner disk regions ($r \sim \SI{0.1}{} - \SI{0.2}{au}$) within $\sim \SI{15}{yrs}$, while adjusting to the new fragmentation limit. 
 }
 \label{Fig_DustDistribution_Final}
\end{figure}

\begin{figure}
\centering
\includegraphics[width=80mm]{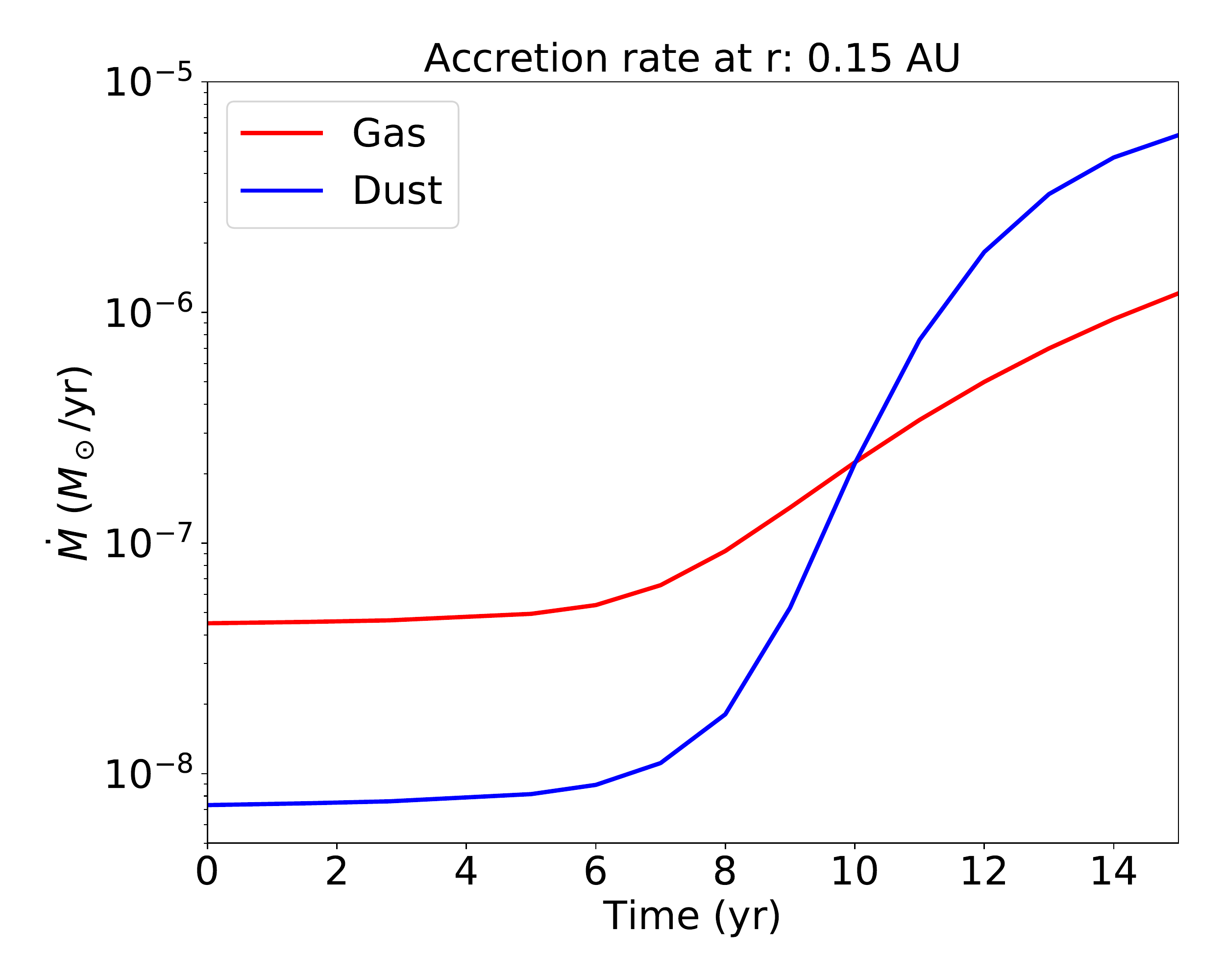}
 \caption{Accretion rate of gas (red) and dust (blue) at $r= \SI{0.15}{au}$ for $\SI{15}{yrs}$ after the dead zone reactivation. The initial gas accretion rate is $\dot{M}_\textrm{g} = \SI{5e-8}{M_\odot/yr}$, in agreement with the observations of RW Aur A, while the dust accretion rate is $\dot{M}_\textrm{d} = \SI{7e-9}{M_\odot/yr}$. Upon reactivation the dust accretion rate increases faster than the gas, eventually surpassing it after \SI{10}{yrs}, and reaching a high value of  $\dot{M}_\textrm{d} = \SI{6e-6}{M_\odot/yr}$.
 }
 \label{Fig_AccretionRates}
\end{figure}

In this section we show the results obtained on the dust and gas dynamics after the \dbquote{dead zone reactivation phase} (Section \ref{sec_Setup_Phase3}), including the final dust distribution, the dust-to-gas ratio at the inner boundary, and accretion rates of dust and gas. 
We also study the impact of the dead zone parameters on the final outcome, to see if these results follow the expected behavior.\\
The gas and dust surface densities before and after the reactivation are shown in \autoref{Fig_FiducialFlushing}. The initial surface density obtained from the first accumulation phase shows a dust-to-gas ratio of $\epsilon = 0.24$ at the dead zone inner edge, and $\epsilon = 0.16$ in the inner disk ($r \lesssim \SI{0.5}{au}$). 
%
After the reactivation ($t > t_r$) the dust accumulated at the dead zone is transported towards the inner regions faster than the gas, reaching the inner boundary of the disk ($r_\textrm{in} \sim \SI{0.1}{} - \SI{0.2}{au}$ \citep{Akeson2005,Eisner2007}) in only $15\ \textrm{years}$.\\
Given that the surface density of dust at the dead zone edge was higher than the gas surface density at the inner regions, this leads to higher concentrations of dust than gas after the reactivation ($\epsilon > 1$). 
This should obviously make the dust dynamically important to the gas motion, however we shall see later in this section that as the particles are too small ($\mathrm{St}<\SI{e-3}{}$), the only impact of dust back-reaction is to slow down the dust and gas evolution. Therefore no instabilities are generated and we can proceed with our analysis.\\
Upon entering the active zone the particles fragment due to the high turbulence and adjust to their fragmentation limit (see \autoref{eq_frag_limit}) in a few collisional timescales $t_\textrm{coll}$, which can be approximated by:
\begin{equation}
t_\textrm{coll} = (n_\textrm{d} \sigma \Delta v_\textrm{turb})^{-1},
\end{equation}
where $n_\textrm{d}$ is the number density of dust particles, $\sigma \approx 4 \pi a^2$ is the collisional cross section, and $\Delta v_\textrm{turb} \approx \sqrt{3 \alpha \St} c_s$ is the turbulent collision speed \citep{Ormel2007}.\\
%
We find that during $\SI{10}{}$ collisional timescales ($t\sim \SI{0.05}{yrs}$) after the dead zone reactivation the dust grains diffuse inward faster than the gas, gaining a head start that leads to high dust-to-gas ratio concentrations.
We attribute this feature to the sudden rise in the turbulence $\alpha$ at the dead zone inner edge, that increases the dust diffusivity and spreads the particles towards the inner regions (see \autoref{eq_dust_diffusion} and \autoref{Fig_DustDistribution_Final}). 
After the dust has adjusted to the new fragmentation limit, it drifts roughly with the viscous velocity of the gas $v_\nu$ towards the inner boundary of the disk. 
The particles reaching the inner boundary reach maximum sizes between $a_\textrm{max} = \SI{10} - \SI{100}{\mu m}$ (see \autoref{Fig_DustDistribution_Final}).\\
The accretion rate (measured at $r = \SI{0.15}{au}$) of both dust and gas increases after the reactivation of the dead zone, as the accumulated material arrives at the inner boundary of the disk (see \autoref{Fig_AccretionRates}).
Before the dead zone reactivation, the gas accretion rate is given by the steady state solution with $\dot{M}_\textrm{g} = \SI{5e-8}{M_\odot/yr}$, similar to the observational value of \citep{Facchini2016, Ingleby2013}, and the dust accretion rate is $\dot{M}_\textrm{d} = \SI{7e-9}{M_\odot/yr}$, this value comes from the dust diffusing into the inner disk during the concentration phase.\\
After the dead zone reactivation the dust concentration moves inwards, and the accretion rate at the inner boundary of the disk becomes dominated by the dust, to the point of surpassing that of the gas. This high supply of solid material, with $\dot{M}_\textrm{d} = \SI{6e-6}{M_\odot/yr}$ could cause hot dust and metallicity features of RW Aur A \citep{Shenavrin2015,Gunther2018}, and provide an ideal environment for the dimmings to occur (see Section \ref{sec_Disc_DimmingContext}). At this point we also note that the accretion rate of gas has increased up to $\dot{M}_\textrm{g} = \SI{e-6}{M_\odot/yr}$.\\
In \autoref{sec_Discussion} we will discuss how the high accretion of solids could cause the dimmings in the context of previous proposed mechanisms (dusty winds, puffed-up inner disk rim, etc), and if we can expect future accretion signatures from the gas. In the following subsections, we study the effect of the simulation parameters on the dust dynamics.

\subsection{Simulation without Dust Concentration}

\begin{figure}
\centering
\includegraphics[width=80mm]{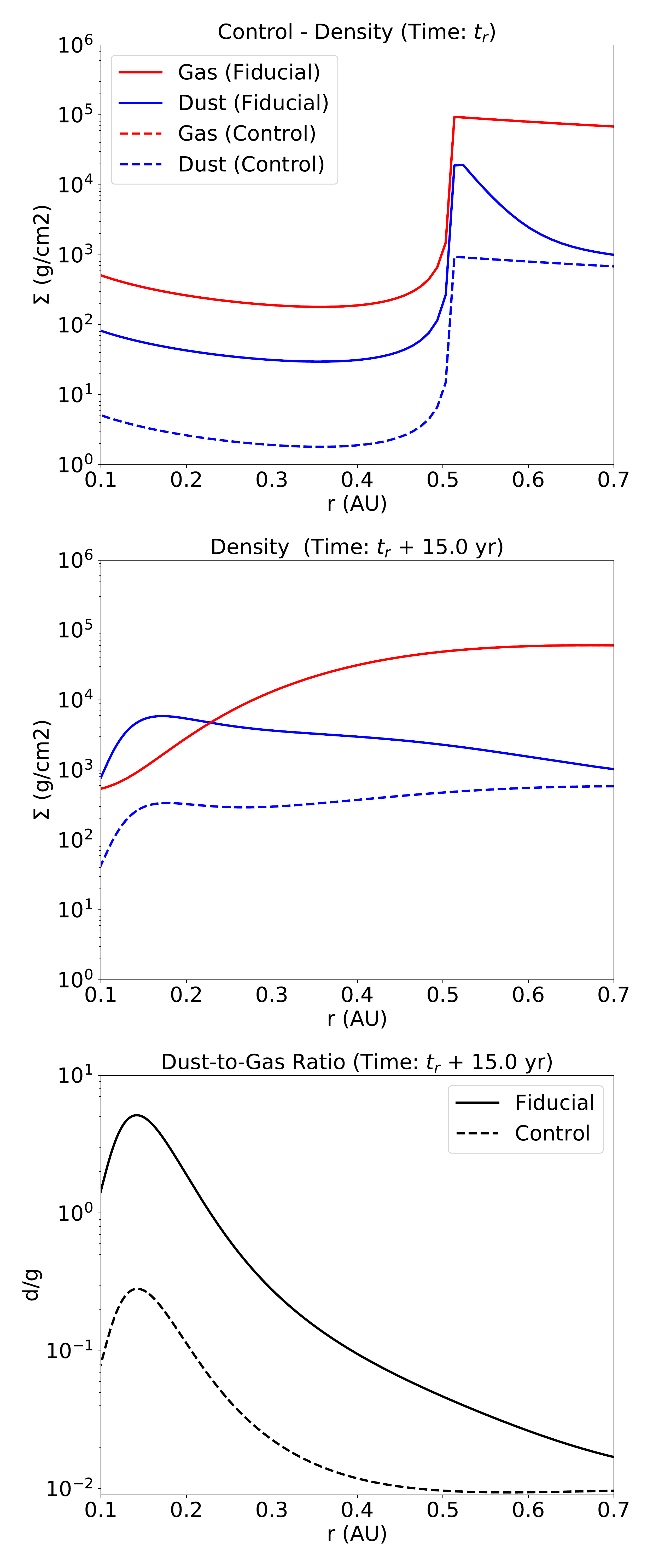}
 \caption{Comparison between the \dbquote{fiducial} simulation ($t_r = \SI{e5}{yrs}$, solid lines) and the \dbquote{control} simulation ($t_r =\SI{0}{yrs}$, dashed lines).
 \textit{Top}: Gas and dust densities at the reactivation time ($t=t_r$). In both cases the gas is in steady state, but in the control simulation the dust did not had the time to accumulate at the inner edge of the dead zone, in this case the reactivation occurs with a uniform dust-to-gas ratio $\epsilon= 0.01$.
 \textit{Mid}: Gas and dust densities after the dead zone reactivation ($t=t_r + \SI{15}{yrs}$). The dust that was at the dead zone in steady state still arrives faster than the gas to the inner boundary of the disk.
 \textit{Bottom}: Dust-to-gas ratio after the reactivation. The maximum dust concentration in the \dbquote{control} simulation is now $\epsilon = 0.28$.
 }
 \label{Fig_ControlFlushing}
\end{figure}

In our model we remained agnostic to the reactivation process of the dead zone, and allowed the dust to accumulate for long enough time to reach concentrations as high as $\epsilon = 0.24$ at its inner edge. 
Depending on the mechanism that reactivates turbulence, the flushing of solid material towards the star may occur earlier with lower dust concentrations, reducing the total accretion of solids. 
To model the limit case in which no dust concentration occurs, we repeat our setup with a control simulation, but now setting the reactivation time to $t_r =  \SI{0}{yrs}$.\\
In \autoref{Fig_ControlFlushing} we show the initial and final state of both simulations. Here the control simulation has an uniform $\epsilon = 0.01$ in the beginning, since no additional dust concentration has occurred. Notice that we still assume that the gas has reached the steady state density profile.\\
After the reactivation the gas and dust excess at the dead zone drift towards the star. As in the fiducial case, the dust that was located at the dead zone inner edge arrives at the inner boundary of the disk before the gas, also in a time of $\sim \SI{15}{yrs}$. 
The only difference is that now the material being accreted has a dust-to-gas ratio of $\epsilon = 0.28$, which is still higher than the initial $\epsilon= 0.01$, although not as extreme as the $\epsilon =5$ found in the fiducial case. Here the dust accretion rate at the inner boundary of the disk ($r \approx \SI{0.15}{au}$) can reach up to $\dot{M}_\textrm{d} = \SI{3e-7}{M_\odot/yr}$.\\
From here we learn that the dust arrival time at the inner boundary does not depend on the amount of solids accumulated at the dead zone inner edge, and that upon reactivation the accreted material will still carry a high concentration of solids.\\
%

\subsection{Simulations for Different Dead Zone Properties}
\begin{figure}
\centering
\includegraphics[width=80mm]{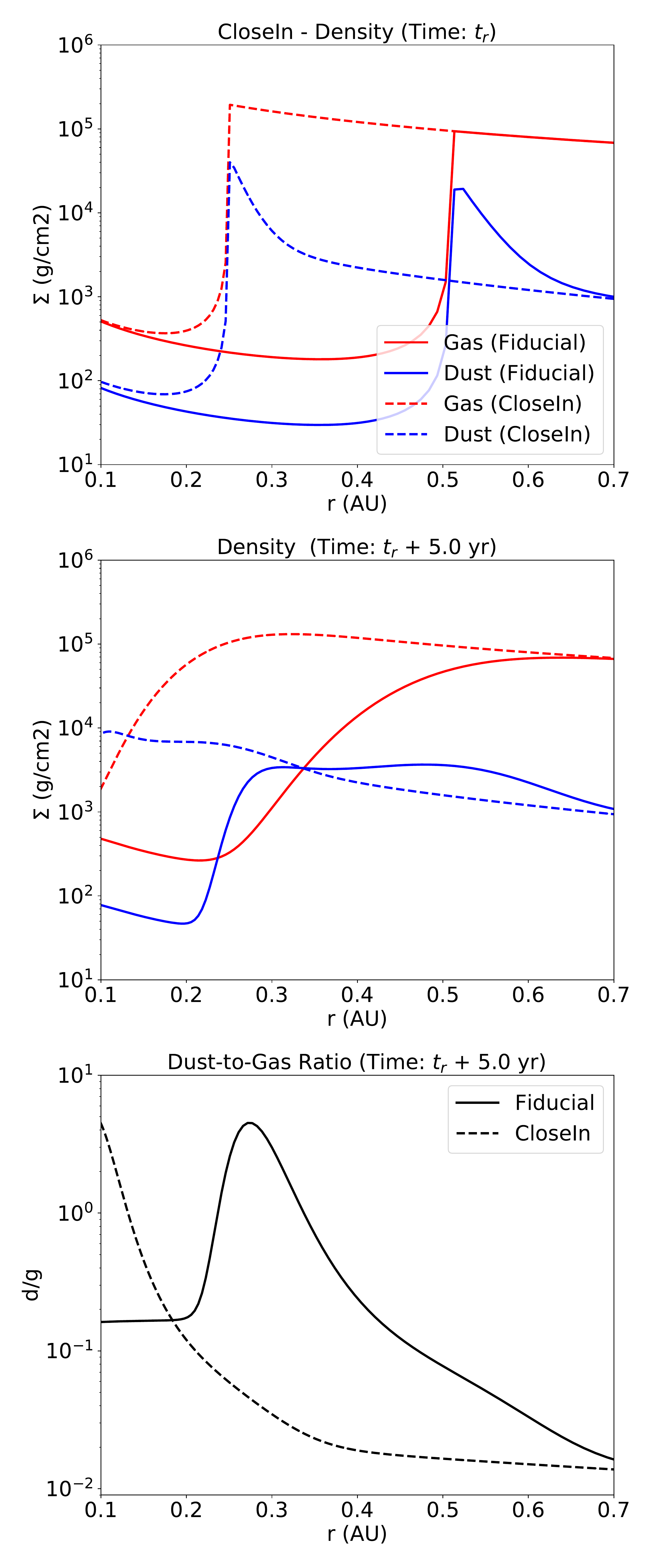}
 \caption{Same as \autoref{Fig_ControlFlushing}, but comparing the \dbquote{fiducial} simulation ($r_1 = \SI{0.51}{au}$) with the \dbquote{closer inner edge} simulation ($r_1 = \SI{0.25}{au}$).
\textit{Top}: Initially, the dead zone is more extended toward the inner boundary of the disk.
\textit{Mid}:  After reactivation, the dust concentration arrives in only $\SI{5}{yrs}$ to the inner edge of the disk, also moving faster than the gas.
 \textit{Bottom}: The maximum dust-to-gas ratio for the \dbquote{closer inner edge} is similar to the \dbquote{fiducial} value with $\epsilon \approx 5$.
 }
 \label{Fig_CloserInnerDZ}
\end{figure}

\begin{figure}
\centering
\includegraphics[width=80mm]{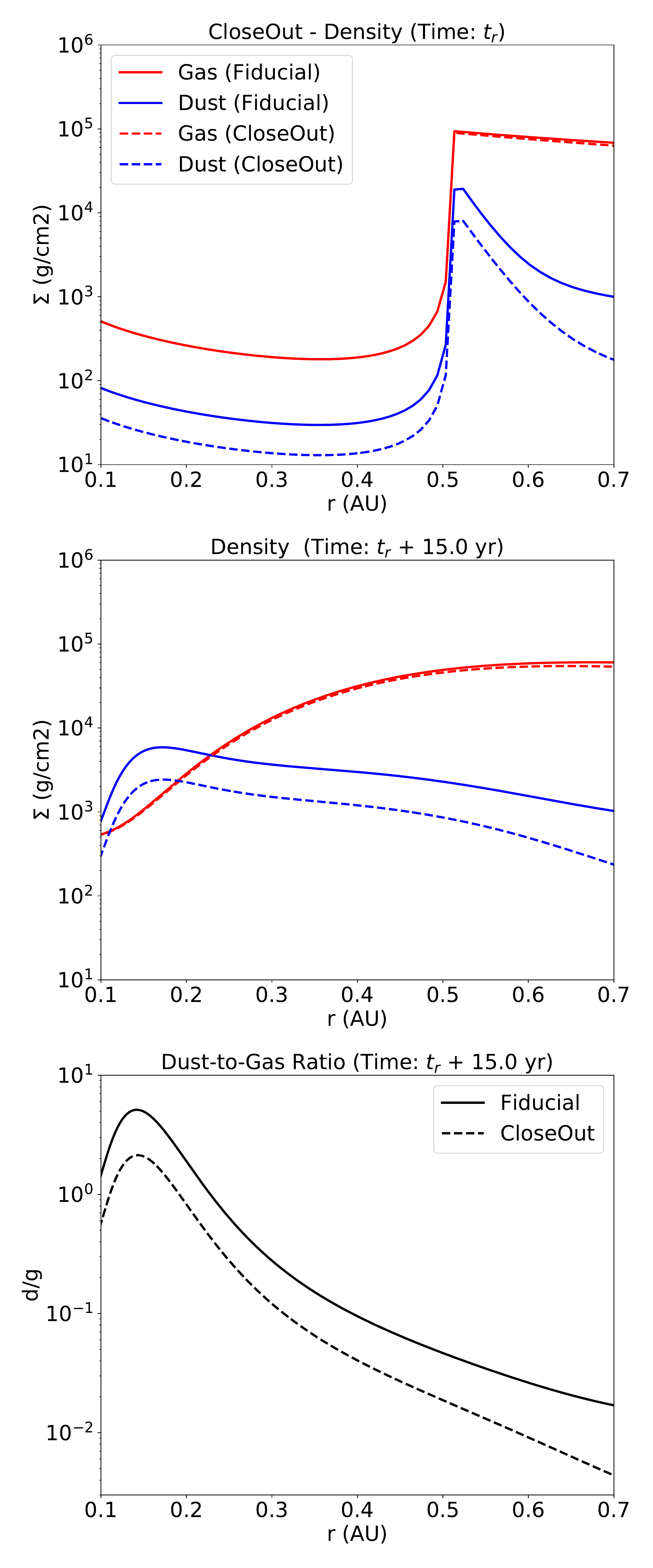}
 \caption{Same as \autoref{Fig_ControlFlushing}, but comparing the \dbquote{fiducial} simulation ($r_2 = \SI{10}{au}$) and the \dbquote{closer outer edge} simulation ($r_2 = \SI{4}{au}$). 
\textit{Top}: Initially the dead zone is smaller and has less material, leading to a lower dust concentration at its inner edge of $\epsilon = 0.1$.
\textit{Mid}: The dust excess once again arrives to the inner boundary of the disk before the gas in $\SI{15}{yrs}$, but in a lower concentration. 
\textit{Bottom}: The maximum dust-to-gas ratio for the \dbquote{closer outer edge} simulation is now $\epsilon\approx 2$.
 }
 \label{Fig_CloserOuterDZ}
\end{figure}

\begin{figure}
\centering
\includegraphics[width=80mm]{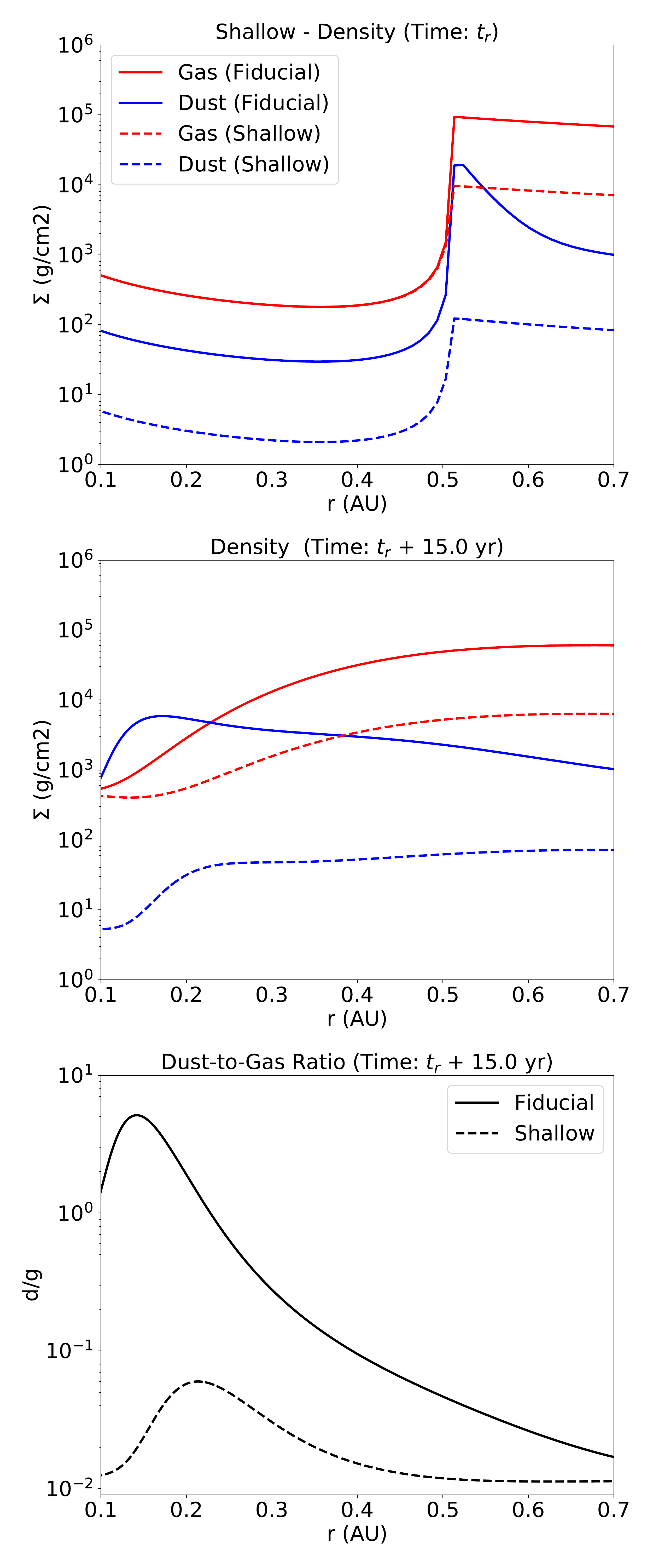}
 \caption{Same as \autoref{Fig_ControlFlushing}, but comparing the \dbquote{fiducial} simulation ($\alpha_\textrm{dead} = \SI{e-4}{}$) and the \dbquote{shallower dead zone} simulation ($\alpha_\textrm{dead} = \SI{e-3}{}$).
 \textit{Top}: There is no notable accumulation of material at the inner edge of the dead zone during the accumulation phase, with $\epsilon = 0.012$ at most. Probably because the particles are too small to get trapped, and diffuse more easily to the inner region.
 \textit{Mid}: After  $\SI{15}{yrs}$ there is only a little dust excess traveling towards the inner region, a small bump can still be appreciated in the dust surface density profile.
 \textit{Bottom}: The final dust-to-gas ratio in the inner boundary is only $\epsilon = 0.06$, just a factor of a few above the original $\epsilon = 0.01$. This is because the lack of more material in the entire dead zone.
 }
 \label{Fig_ShallowerDZ}
\end{figure}

The dead zone shape is parametrized by its edges $r_1$ and $r_2$, and the turbulence parameter $\alpha_\textrm{dead}$ following \autoref{eq_alpha_profile}, and altering these parameters also changes arrival time and dust-to-gas ratio of the accreted material.\\
By shifting the inner edge of the dead zone to smaller radii ($r_1 = \SI{0.51}{au} \rightarrow \SI{0.25}{au}$) the dust concentration during the first phase will also be located closer to the inner boundary of the disk (see \autoref{Fig_CloserInnerDZ}). Now it only takes the dust between $\SI{3}{} - \SI{5}{yrs}$ to reach the inner boundary of the protoplanetary disk. Therefore the inner edge of the dead zone regulates the time required between the reactivation and the accretion of the dusty material.\\
A dead zone with a closer outer boundary ($r_2 = \SI{10}{au} \rightarrow \SI{4}{au} $) will be smaller and concentrate less dust at its inner boundary (see \autoref{Fig_CloserOuterDZ}). This also reduces the total amount of solid material that it is accreted towards the star, although this is still a considerable amount with a dust-to-gas ratio of $\epsilon = 2.1$.\\
Finally, the most significant parameter is the turbulence $\alpha_\textrm{dead}$ of the dead zone. Our fiducial simulation considered an $\alpha_\textrm{dead} = \SI{e-4}{}$, which in contrast with the active zone $\alpha_\textrm{active} = \SI{e-1}{}$ leads to an accumulation of material in the dead zone with a factor of 1000 relative to the steady state of a fully active disk, this of course favors the accretion of massive amounts of gas and solids upon reactivation.
By taking a shallower dead zone ($\alpha_\textrm{dead} = \SI{e-4}{} \rightarrow \SI{e-3}{}$) there is less gas and dust accumulated, so upon reactivation the flushing of material is slower by a factor of a few (see \autoref{Fig_ShallowerDZ}). We also find that for $\alpha_\textrm{dead} = \SI{e-3}{}$ there is no significant concentration at the dead zone inner edge, after $\SI{e5}{yrs}$ the dust-to-gas ratio rises only up to $\epsilon = 0.012$. This happens because the higher turbulence lowers the fragmentation limit (see \autoref{eq_frag_limit}), producing particles that drift slower towards the pressure maximum (which now is also shallower).\\
The final results is that the accreted material after the dead zone reactivation only has a dust-to-gas ratio of $\epsilon = 0.06$.
Therefore we find that the turbulence parameter $\alpha_\textrm{dead}$ of the dead zone is the main determinant of the total amount of material being accreted, and that deeper dead zones are necessary to produce the dust dominated accretion seen in our fiducial simulation.

\subsection{Simulation with Dust Back-reaction}\label{sec:backreaction}

\begin{figure}
\centering
\includegraphics[width=80mm]{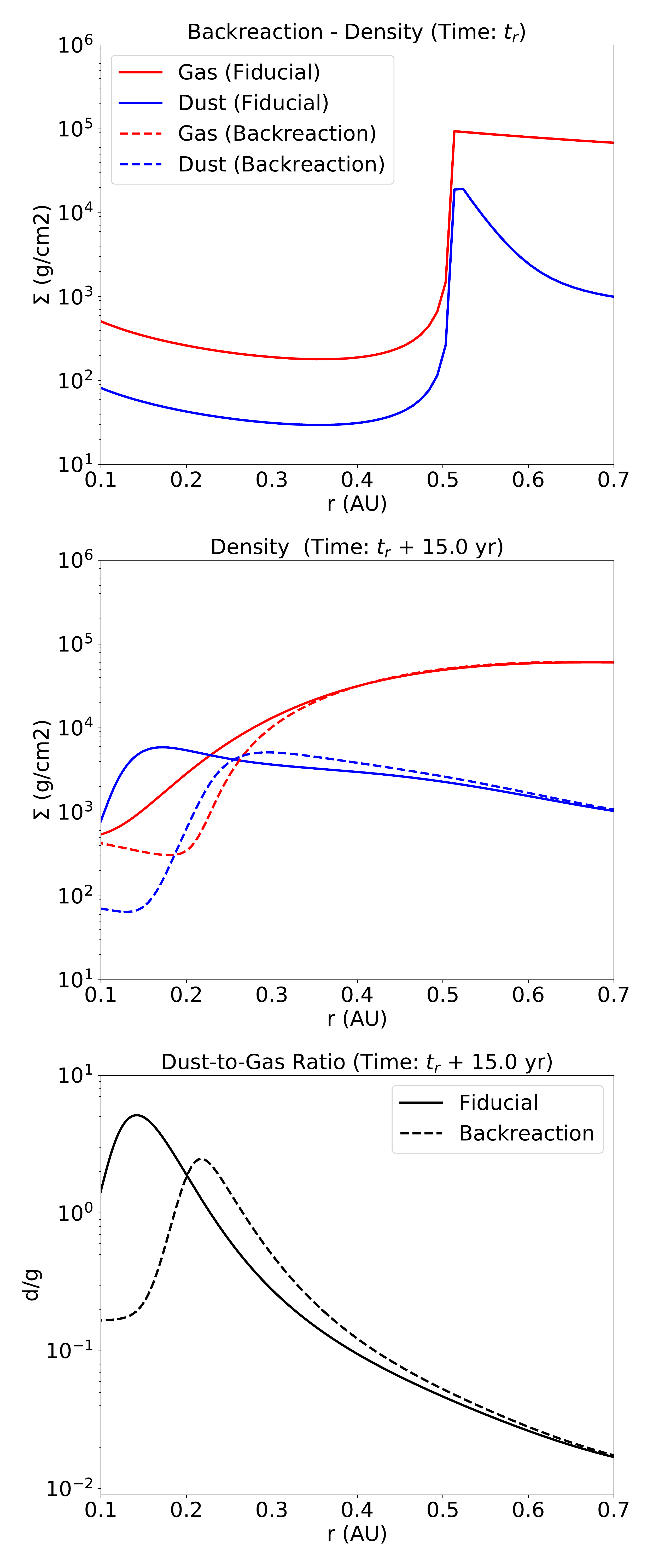}
 \caption{Same as \autoref{Fig_ControlFlushing}, but comparing the \dbquote{fiducial} simulation and the \dbquote{back-reaction} simulation.
 \textit{Top}: Both simulations start at the reactivation time with the same initial conditions.  
 \textit{Mid}: When back-reaction is considered, the evolution of the dust and gas component is slower than in the fiducial case. This is because the high dust concentrations slow down the viscous evolution of the gas, which in turn also slows down the drifting of the dust towards the inner disk.
\textit{Bottom}: Both the simulation with and without back-reaction present a high concentration of dust in the accreted material, yet in the case with back-reaction the bulk of dust reaches a radii of only $r = \SI{0.22}{au}$ in \SI{15}{years}, while the dust in the fiducial simulation is already at $r = \SI{0.15}{au}$.
 }
 \label{Fig_Backreaction}
\end{figure}

In all our results until this point we have neglected the back-reaction of dust to the gas, however for the dust-to-gas ratios presented during the dead zone reactivation ($\epsilon \gtrsim 1$) this effect should be relevant. In this section we study its impact to see if our previous results remain valid.\\
First, we should mention that in this setup we only consider the back-reaction in the \dbquote{reactivation phase}, and assume that the dust will still accumulate at the inner edge of the dead zone even if back-reactions are considered. This is still justified since our particles are too small ($\mathrm{St} < \SI{e-3}{}$) to cause any perturbation beyond slowing down the concentration process and we can be infer it also by studying the single dust species scenario described in the Appendix \ref{sec_Appendix_SingleBR}. 
Studies of \citet{Onishi2017} also showed that the back-reaction still allows the dust to accumulate at pressure maxima, and that the dust traps do not self-destruct by this effect when taking into account the vertical distribution of solids.\\
For the reactivation phase we implement the gas velocities as described by equations \autoref{eq_backreaction_vr} and \autoref{eq_backreaction_vtheta}. 
In the radial direction, the gas velocity now consist of two terms modulated by the back-reaction coefficients $0<A,B<1$, the term $A v_\nu$ is slowing down the viscous evolution of the gas respect to the default value $v_\nu$, and the term $2 B \eta v_\textrm{K}$ is pushing the gas in the direction opposite to the pressure gradient.\\
Since the vertical distribution of particles is not exactly the same as the gas, the effect of the back-reaction is also not uniform in the vertical direction, the importance of this point is shown in \citet{Dipierro2018,Onishi2017}. 
To account for the vertical effect of back-reactions in our 1D simulations we take the vertically averaged velocity for the dust and gas, weighted by the mass density to conserve the total flux. The details for this implementation can also be found in the Appendix \ref{sec_Appendix_VerticalBR}.\\
In \autoref{Fig_Backreaction} we show a comparison between the \dbquote{fiducial} and \dbquote{back-reaction} simulations \SI{15}{years} after the dead zone reactivation. 
When the back-reaction is considered the most notable effect is the slowing down of the accretion of material  by a factor of $\sim 2$.
While the dust particles in the \dbquote{fiducial} simulation take $15\ \textrm{years}$ to reach the inner boundary of the disk, in the \dbquote{back-reaction} simulation they need \SI{30}{years} instead.\\
The reason why no further effects are observed is that the high turbulence causes fragmentation of the particles to smaller sizes (as seen in \autoref{Fig_DustDistribution_Final}), where they are unable to \dbquote{push} the gas backwards (i.e. the term $B \rightarrow 0$), and are only able to reduce the gas viscous velocity by a factor of $A \leq 1$.\\
Another effect of the back-reaction is that the maximum dust-to-gas ratio has also decreased to a value of $\epsilon=2.5$ because of the dust and gas redistribution by their coupled interaction (the dust concentration does not increase with time to the fiducial $\epsilon=5$ value).
This means that in all our previous results we should consider that the accreted material will take more time (a factor $\sim 2$, if the dust-to-gas ratio is high enough) to reach the inner boundary of the protoplanetary disk, and as consequence that the accretion rate will also be reduced.\\
An order of magnitude estimate for the back-reaction coefficients can be obtained by approximating the particle distribution to a single size population (see  Appendix \ref{sec_Appendix_SingleBR}), in which $A \approx (\epsilon +1 )^{-1}$ and $B \approx \mathrm{St}\ \epsilon \ (\epsilon +1)^{-2}$, however to have an overall estimate for the entire disk the dust-to-gas distribution should be taken into account.
%
%
\section{Discussion} \label{sec_Discussion}

We have seen that through the reactivation of a dead zone, located in the inner regions of RW Aur A circumstellar disk, large amounts of dust can be flushed towards the star in timescales that can go between  $\SI{5}{} - \SI{30}{years}$. In this section we compare our results on dust accretion with the properties of the dimmings, speculate upon the future accretion signatures of RW Aur, and discuss which ingredients of our model may be improved to better match the observations.

\subsection{The Fast Accretion Mechanism in the Context of RW Aur A Dimmings} \label{sec_Disc_DimmingContext}
Multiple observations during the dimmings of  RW Aur A reveal the presence of large amounts of dust at small radii, in the line of sight, and in the material accreted by the star (see Section \ref{sec_Intro_Observations}). 
Our model with dead zone reactivation provides a way to increase the dust at the inner rim of the protoplanetary disk by 2 orders of magnitude from its original value (from $\Sigma_\textrm{d} = \SI{50}{g/cm^2}$ to $\SI{5e3}{g/cm^2}$ at $\SI{0.15}{au}$). Although this alone does not explain the dimmings, as a significant amount of material still needs to be moved towards the line of sight, it relaxes the conditions of other mechanisms that can do so.\\
In the case that the dimmings are caused by a dusty wind coming from the inner regions \citep{Petrov2015,Shenavrin2015}, the dead zone reactivation increases the amount of dust entangled with the gas, and then the particles would be dragged into the line of sight by the wind if they are small enough to be coupled to the gas. \\
Similarly, if the dimmings are caused by a puffed up inner rim of the disk  \citep{Facchini2016,Gunther2018}, the reactivation of the dead zone not only increases the amount of solid material in the line of sight (since our particles are small they should be well mixed with the gas in the vertical direction), but also can increase the scale height of these regions with the rise in temperature due to accretion heating. 
If our model is correct, in the following decade(s) the gas accretion rate should also increase by more than one order of magnitude, up to $\dot{M}_\textrm{g} = \SI{e-6}{M_\odot/yr}$, as the gas excess from the dead zones adjusts to a new steady state.
%
%
\subsection{A Single Reactivation Event, or Multiple Short Reactivation Spikes?} \label{sec_Disc_ReactivationMechanism}
The dimmings of RW Aur A last from a few months up to two years \citep[see the list in][]{Rodriguez2018}, repeatedly moving from the dimmed state to the bright state. The accretion process described in our model last from several years to a few decades (depending on the simulation parameters), which is clearly longer than the typical dimming duration, and without presenting any decrease in the dust accretion rate over the process. Yet, we can think of a few ways to reconcile the timescales of our simulations with the dimmings.\\
The first possibility is that the reactivation of the dead zone, and the subsequent increment in the accretion rate (see \autoref{Fig_AccretionRates}), can also raise the temperature of the inner disk through viscous heating, resulting in a puffed-up inner rim.
In this case, the dimmings will occur in the time required to drag the dusty material into the line of sight as the scale height of the inner disk increases, and will finish once the material settles back down or is completely accreted.\\
In a similar way, if the dead zone reactivation is accompanied by stronger stellar winds, these will determine the timescale of the dimmings within the accretion event described by our model.\\
Also, the accretion process described in our model does not need to increase smoothly, and can present variability in shorter timescales.
Instabilities that are not resolved by our model during the accretion process may produce a bumpy surface density profile in dust and gas, for example through the ring instability \citep{Wunsch2005} for dead zones in layered disks. 
In this case the dimming events would correspond to the local maximums in the accretion of dust.\\
Another possibility is that the dead zone reactivation is not instantaneous. In the case that only inner edge becomes active, a fraction of the accumulated dust will start drifting towards the star, leaving most of the gaseous and solid material still trapped in the dead zone. 
If multiple reactivation events like this take place, the corresponding dust excess will also arrive in intervals, with a frequency depending on the reactivation mechanism. In this case the dimmings would start when a spike of dust accretion arrives to the inner edge, and end once it decreases back to its steady state value. To improve our model we would need to resolve the thermal and gravitational instabilities that can reactivate the dead zone \citep{Martin2011}.\\
Finally, azimuthal asymmetries in the inner disk (such as vortices) may also add variability to the accretion process, but these are not considered by our model.\\
A further monitoring of the metallicity of the accreted material would help understand the nature of the dust accretion process and the mechanism that drags the dust into the line of sight. 
If the dust accretion continues delivering material to the star, independently of the dimmings, then the metallicity should remain high even after the luminosity returns to the bright state. Otherwise, if the rise in metallicity keeps correlating with the dimmings \citep[as in][]{Gunther2018}, the dimmings could be an outcome of the sudden increase in the dust accretion (although of course, correlation does not imply causality).

\subsection{Validity of the Dead Zone Model} \label{sec_Disc_DeadZone_Model}
Our results showed that the exact values of dust accretion rate and timescales depend sensitively on the parameters used for the dead zone. A dead zone inner edge closer to the inner boundary of the disk reduces the timescale of the process, a closer outer edge reduces the total mass of the dead zone, and the turbulence parameter $\alpha_\textrm{dead}$ and the reactivation time determine the amount of dust that can be trapped and flushed towards the star. 
In addition, the shape of the dead zone profile also affects dust and gas surface density profiles, and a proper modeling of the gas turbulence would be required to obtain their final distributions after the dead zone reactivation.\\
With this amount of free parameters, our simulations provide more a qualitative scenario than quantitative predictions. Further constrains are necessary to determine how relevant the reactivation of a dead zone can be for RW Aur A dimmings. 
The first step is of course to find the mechanism that puts the required amounts of dust in the line of sight, and obtain an estimate of the required dust surface density at the inner disk (not only in the line of sight) to produce this phenomena.\\
In parallel, any signature of enhanced gas accretion rate in the following years would speak in favor of the dead zone reactivation mechanism as one of the drivers of the dimmings.\\ 
Additionally, constrains on the properties of the inner disk would limit the parameter space described. 
In particular, the mass in the inner \SI{10}{au} of RW Aur A disk would be useful to constrain the outer edge and turbulence parameter of the dead zone, and with them the amount of material that can be thrown to the star.

\subsubsection{The Dead Zone as an Accretion Reservoir}
One point that speaks in favor of our model is its potential to sustain the large accretion rates of RW Aur for extended periods of time. 
Considering only the observed values for the disk mass and the accretion rate, the maximum lifetime of RW Aur would be of $M_\textrm{g}/\dot{M}_\textrm{g} \sim \SI{e4}{}$ - $\SI{e5}{yrs}$ which is too short for a T Tauri star.\\
\cite{Rosotti2017} defined the dimensionless accretion parameter:
\begin{equation}
\eta_\textrm{acc} =  \tau_* \dot{M}/M_\textrm{disk},
\end{equation}
with $\tau_*$ the age of the star, that indicates if the properties of a disk are consistent the steady state accretion, in which case it follows $\eta_\textrm{acc} \lesssim 1$. The RW Aur A is around \SI{5}{Myr} old \citep{Ghez1997} and presents an accretion parameter of $\eta_\textrm{acc} \approx 60$, indicating either that the disk is not in steady state, or that the observed disk mass is underestimated.\\
In our model, the dead zone provides a reservoir of material able to sustain the detected accretion rates for around {2}{Myrs} (considering the parameters of our fiducial simulation), which is close to the estimated age of the star. 
At the same time, at high densities the dusty material would be optically thick, remaining hidden from the mm observations used to measure the disk mass.
\subsubsection{Another Free Parameter for Dust Growth?} \label{sec_Disc_TwoAlphas}
In our model we considered that the turbulent $\alpha$ that limits dust growth by fragmentation (\autoref{eq_frag_limit}), is the same that drives the viscous evolution of the gas (\autoref{eq_alpha_visc}), yet recent models explore the case with two independent $\alpha$ values for each process \citep[e.g.,][]{Carrera2017}.
For our model, using a single $\alpha$ value means that particles reach bigger sizes while they remain in the dead zone, and fragment to smaller sizes in the active region.\\
Allowing two independent alpha values would allow the formation of large particles in the active region.  These larger particles drift faster, but also exert a stronger back-reaction on the gas. The \dbquote{pushing} back-reaction coefficient $B$ is roughly proportional to the particle size (see Appendix \ref{sec_Appendix_SingleBR}), and at the high dust-to-gas ratios found during the dead zone reactivation, it could be strong enough to generate density bumps in the gas, or even trigger the streaming instability for particles with large enough Stokes number \citep{Youdin2005}. \\

\section{Summary}\label{sec_Summary}
In this work we studied a new mechanism that can increase the concentration of solids in the inner regions of a protoplanetary disk in timescales of $\sim \SI{10}{years}$, through the reactivation of a dead zone.\\
This study was motivated by the recent dimmings of RW Aur A, which present a high concentration dust in the line of sight \citep{Antipin2015,Schneider2015}, and an increased emission from hot grains coming from the inner regions of the protoplanetary disk\citep{Shenavrin2015}, and subsequently observed super-solar metallicity of the accreted material\citep{Gunther2018}.\\
Using 1D simulations to model the circumstellar disk of RW Aur A,
we find that the dust grains accumulate at the inner edge of the dead zone, which acts as a dust trap, reaching concentrations of $\epsilon \approx 0.25$. When the turbulence in this region is reactivated, the excess of gas and dust is released from the dead zone and advected towards the star. By effect of dust diffusion and gas drag, the dust component can reach the inner boundary of the protoplanetary disk before the gas component, producing high dust concentrations of $\epsilon \approx 5$.\\
The accretion rate of solids increases from $\dot{M}_\textrm{d} = \SI{7e-9}{M_\odot/yr}$ to $\SI{6e-6}{M_\odot/yr}$ in only $\SI{15}{years}$. This scenario can provide an ideal environment for other mechanisms, such as stellar winds \citep{Petrov2015,Shenavrin2015} or a puffed up inner rim \citep[e.g.,][]{Facchini2016,Gunther2018}, to transport the required amount of solid  material into the line of sight and cause the dimmings, although further studies are required to link the surface density at the midplane with the measured dust concentrations.\\
Additionally, our simulations predict that in the following decade(s) the gas accretion rate should also rise by an order of magnitude, from $\dot{M}_\textrm{g} = \SI{5e-8}{M_\odot/yr}$ to $\SI{e-6}{M_\odot/yr}$  if the dead zone reactivation is the mechanism transporting dust towards the disk inner region.

\section*{Acknowledgments}
We would like to thank S. Facchini, and also the anonymous referee, for their useful comments that improved the extent of this work.
M. G., T. B., and S. M. S. acknowledge funding from the European Research Council (ERC) under the European Union’s Horizon 2020 research and innovation programme under grant agreement No 714769 and funding by the Deutsche Forschungsgemeinschaft (DFG, German Research Foundation) Ref no. FOR 2634/1.
HMG was supported by the National Aeronautics and Space Administration through Chandra Award Number GO6-17021X issued by the Chandra X-ray Observatory Center, which is operated by the Smithsonian Astrophysical Observatory for and on behalf of the National Aeronautics Space Administration under contract NAS8-03060.

\appendix

\section{Details of Dust Back-reaction}
\label{sec_Appendix_BackReaction}
The gas and dust velocities incorporating the back-reaction of the dust into the gas is obtained from the momentum conservation equations shown in \citet{Nakagawa1986}, and considering the force exerted by the multiple species of dust as in \citep{Tanaka2005}.
In this case the gas feels the drag force of multiple species, the pressure force, the viscous force, and the stellar gravity, while the dust only feels the stellar gravity and the drag force from the gas.\\
The momentum equations for the gas and dust then are respectively:
\begin{equation} \label{eq_momentum_conservation_gas}
\frac{\stdiff{\mathbf{v_\textrm{g}}} }{\stdiff{t}} =  -\int \frac{(\mathbf{v_\textrm{g}} - \mathbf{v_\textrm{d}})}{t_\textrm{stop}} \frac{\rho_\textrm{d}(m)}{\rho_\textrm{g}} \stdiff{m} - \frac{G M_*}{r^2}\mathbf{\hat{r}} - \frac{1}{\rho_\textrm{g}}\frac{\partial P}{\partial r} \hat{r} + f_\nu \mathbf{\hat{\theta}},
\end{equation}
\begin{equation} \label{eq_momentum_conservation_dust}
\frac{\stdiff{\mathbf{v_\textrm{d}}} }{\stdiff{t}} =  - \frac{(\mathbf{v_\textrm{d}} - \mathbf{v_\textrm{g}})}{t_\textrm{stop}} - \frac{G M_*}{r^2} \mathbf{\hat{r}},
\end{equation}
where the viscous force contribution in the azimuthal direction can be conveniently written as $f_\nu = \Omega_K v_\nu / 2$.\\
Solving this system of equations for the steady state gives the radial and azimuthal velocities for gas as shown in \autoref{eq_backreaction_vr}, \autoref{eq_backreaction_vtheta}, and the radial velocity for the dust given by \autoref{eq_dust_velocity}.\\
The back-reaction coefficients $A$, $B$ are defined as:\\
\begin{equation} \label{eq_backreaction_A}
A = \frac{X + 1}{Y^2 +(X+1)^2},
\end{equation}
\begin{equation} \label{eq_backreaction_B}
B = \frac{Y}{Y^2 +(X+1)^2},
\end{equation}
with $X$ and $Y$ following the notation of \citet{Okuzumi2012}:
\begin{equation} \label{eq_backreaction_X}
X = \int \frac{1}{1+\mathrm{St}^2} \frac{\rho_\textrm{d}(m)}{\rho_\textrm{g}} \stdiff{m},
\end{equation}
\begin{equation} \label{eq_backreaction_Y}
Y = \int \frac{\mathrm{St}}{1+\mathrm{St}^2} \frac{\rho_\textrm{d}(m)}{\rho_\textrm{g}} \stdiff{m}.
\end{equation}
The $X,Y$ integrals come from the contribution of the multiple dust species at the momentum conservation equations described by \citet{Tanaka2005}. Here $\rho_\textrm{d}(m)$ is the dust volume density per mass, and the dust-to-gas ratio at each mass bin $\epsilon(m) = (\rho_\textrm{d}(m)/\rho_\textrm{g})dm$ determines the contribution of each particle species to the final result.\\
To summarize the effect of back-reactions on the gas, we can understand the coefficient $A$ as a \dbquote{smoothing} factor and the coefficient $B$ as a \dbquote{pushing} factor.
The \dbquote{smoothing} factor $A$ reduces the effect of the viscous force in the radial velocity (slowing down the accretion), and reduces the effect of the pressure gradient in the azimuthal velocity (making the gas orbital velocity more keplerian). The \dbquote{pushing} factor $B$ on the other hand, pushes the gas in the direction opposite to the pressure gradient in the radial direction (with the term $2 B \eta v_K$), and slows down the orbital velocity (with the term $B v_\nu /2$).
Which term dominates in the radial evolution of dust and gas will also depend on the magnitude of the pressure gradient and the viscosity \citep{Dipierro2018}.

\subsection{Single Species Analysis} \label{sec_Appendix_SingleBR}
The expressions for $A$ and $B$ are difficult to study by eye, but we can consider the case with a single species of particles to simplify subsequent analysis. In this case the integrals $X$ and $Y$ become:
\begin{equation} \label{eq_backreaction_X_single}
X_\textrm{single} = \frac{1}{1+\mathrm{St}^2} \epsilon,
\end{equation}
\begin{equation} \label{eq_backreaction_Y_single}
Y_\textrm{single} = \frac{\mathrm{St}}{1+\mathrm{St}^2} \epsilon.
\end{equation}
The back-reaction coefficients now have a simple expression that only depends on the Stokes number and the dust-to-gas ratio:
\begin{equation} \label{eq_backreaction_A_single}
A_\textrm{single} = \frac{\epsilon + 1 + \mathrm{St}^2}{(\epsilon + 1)^2 + \mathrm{St}^2},
\end{equation}
\begin{equation} \label{eq_backreaction_single}
B_\textrm{single} = \frac{\epsilon \mathrm{St}}{(\epsilon + 1)^2 + \mathrm{St}^2}.
\end{equation}
From these equations we can start noticing some interesting values for $A$ and $B$:
\begin{itemize}
\item $0 < A,B < 1$
\item The limit without particles $\epsilon \rightarrow 0$, recovers the traditional gas velocities:
	\begin{itemize}
      \item $A \rightarrow 1$
      \item $B \rightarrow 0$
	\end{itemize}
\item The limit with small particles, assuming $\mathrm{St}<<1$ gives:
    \begin{itemize}
      \item $A \approx (\epsilon +1 )^{-1}$
      \item $B \approx \mathrm{St}\ \epsilon \ (\epsilon +1)^{-2}$
    \end{itemize}
\end{itemize}
The last case is applicable to our simulations, since we have high dust-to-gas ratios of small particles only, in this case we reach the limit of $B\approx 0$ and $A\approx (1+\epsilon)^{-1}$, and therefore we only care about the slowing effect of the particles on the gas.\\

\subsection{Considering the Vertical Structure} \label{sec_Appendix_VerticalBR}
The dust-to-gas ratio is not necessarily constant in the vertical direction. As shown by \citet{Okuzumi2012,Dipierro2018} this might have an impact in the effects of the back-reaction, since the velocities of the gas and dust will experience a different force in the different layers of the disk.\\
We assume that both the gas and dust densities are distributed as a Gaussian in the vertical direction:
\begin{equation} \label{eq_vertical_density_gas}
\rho_\textrm{g}(z) = \frac{\Sigma_\textrm{g}}{\sqrt{2\pi} H_\textrm{g}} \exp(-\frac{z^2}{2 H^2_\textrm{g}}),
\end{equation}
\begin{equation} \label{eq_vertical_density_dust}
\rho_\textrm{d}(m,z) \stdiff{m} = \frac{\Sigma_\textrm{d}(m)}{\sqrt{2\pi} H_\textrm{d}(m)} \exp(-\frac{z^2}{2 H^2_\textrm{d}(m)}).
\end{equation}
Here $\rho_\textrm{d}(m,z) dm$ is the volume density of the particles in the mass bin $m$ at a height $z$, and $\Sigma_\textrm{d}(m)$ is the surface density of particles at the mass bin $m$. $H_\textrm{g}$ and $H_\textrm{d}(m)$ are the scale heights of the gas and the dust particles with mass $m$, respectively defined as:
\begin{equation} \label{eq_Scaleheight_gas}
H_\textrm{g} = \frac{c_s}{\Omega_K},
\end{equation}
\begin{equation}\label{eq_scaleheight_dust}
H_\textrm{d}(m) = H_\textrm{g} \cdot \min(1, \sqrt{\frac{\alpha}{\min(\mathrm{St},1/2) (1+\mathrm{St}^2)  }}),
\end{equation}
the last equation coming from \citet{Birnstiel2010}, that tells us that bigger particles are more concentrated towards the midplane, while small particles are distributed like the gas.\\
With the vertical distribution of gas and dust we can obtain the mass weighted average radial velocity $\bar{v}_\textrm{g,d}$, defined to conserve the radial mass flux, as:
\begin{equation} \label{eq_vertical_average_velocity}
\Sigma_\textrm{g,d} \bar{v}_\textrm{g,d} = \int_{-\infty}^{+\infty} \rho_\textrm{g,d}(z) v_\textrm{g,d}(z) \stdiff{z}.
\end{equation}
Now the densities, dust-to-gas ratio and Stokes number of the particles can be defined in the vertical direction, allowing us to have the vertical distribution of back-reaction coefficients and radial velocities.\\ 
In terms of implementation, we perform the integral \autoref{eq_vertical_average_velocity} over $n_z=50$ logarithmically spaced grid cells locally defined between $10^{-3} - 4 H_\textrm{g}$.
Our only assumption to simplify the numeric calculation is that the viscous velocity $v_\nu$ and the sub-keplerian velocity $\eta v_K$ are constant over the vertical direction.
%
\section{Higher Resolution Test}
\begin{figure}
\centering
\includegraphics[width=80mm]{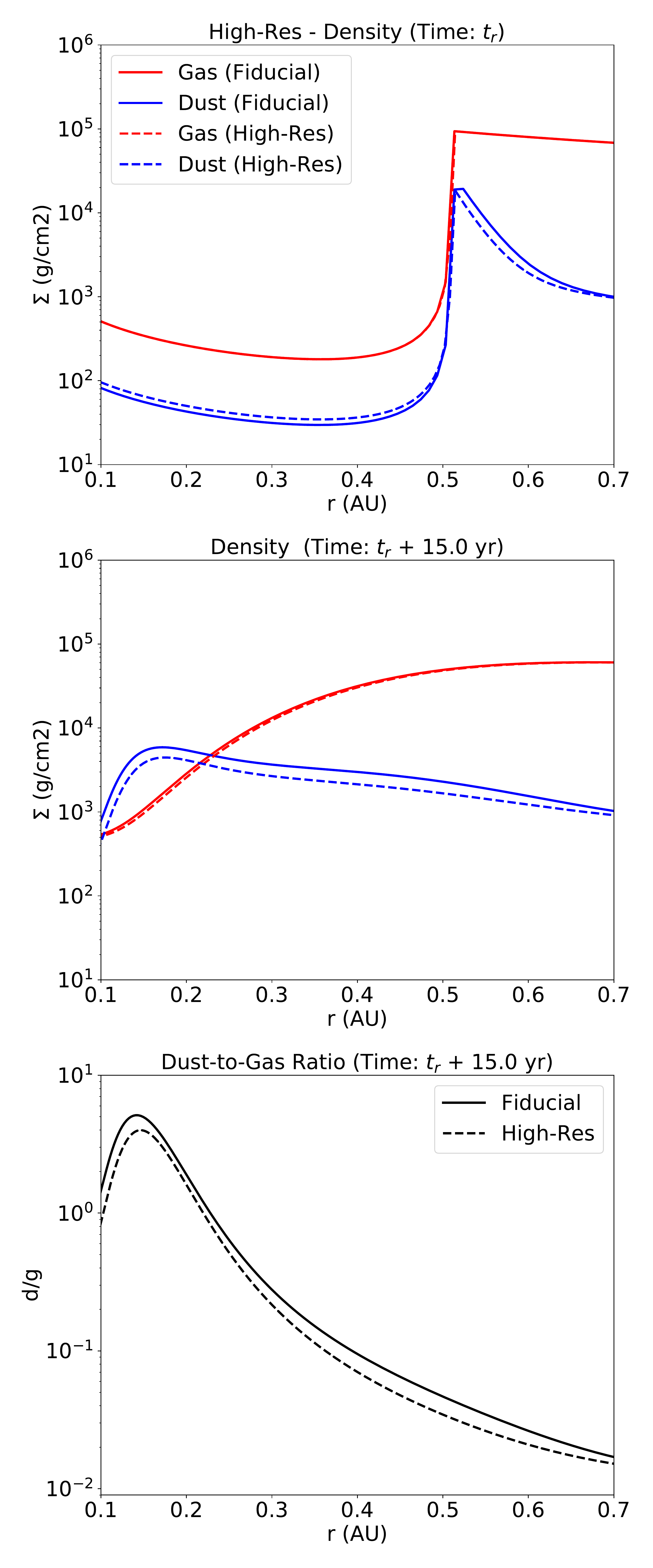}
 \caption{Comparison between the \dbquote{fiducial} (solid line) simulation and the \dbquote{high resolution} (dashed line) simulation.
 \textit{Top}: At higher resolutions, there is less dust accumulated at the dead zone inner edge, due to more effective diffusion of solids towards the inner regions in the dust concentration phase.
 \textit{Mid}: After \SI{15}{yrs} of evolution, the gas profile remains practically identical with the increased resolution, the dust profile is slightly lower at all radii in the high resolution case.  
 \textit{Bottom}: The final dust-to-gas ratio in the high resolution case is slightly reduced, the maximum dust concentration now has a value of $\epsilon \approx 4$. No further changes are observed.
 }
 \label{Fig_HighResComparison}
\end{figure}

To validate our results we perform an additional simulation using the same parameters as in the fiducial set-up, but with an increased radial resolution for the different phases.\\
In the Phase 1 of dust concentration (Section \ref{sec_Setup_Phase1}) we increase to $n_r = 1000$ radial grid cells between $r = \SI{0.01} - \SI{100}{au}$. In the Phase 2 and 3 of build up and dead zone reactivation (Section \ref{sec_Setup_Phase2}, \ref{sec_Setup_Phase3}) we use:
\begin{itemize}
\item 40 linear-spaced grid cells at $r = \SI{0.05}{} - \SI{0.09}{au}$,
\item 200 logarithmic-spaced grid cells at $r = \SI{0.09}{} - \SI{1.0}{au}$,
\item 30 logarithmic-spaced grid cells at $r = \SI{1.0}{} - \SI{5.0}{au}$,
\end{itemize}
and keep the mass grid to $n_m = 141$ logarithmically spaced grid cells.\\
In \autoref{Fig_HighResComparison} we see that the evolution of gas remains the same independent of the resolution used. The final dust surface density profile and the dust-to-gas ratio are slightly lower for the higher resolution simulation over the entire inner region, the maximum concentration is reduced from $\epsilon = 5$ to $\epsilon=4$.\\
This decrease in the dust surface density apparently happens because, during the dust concentration phase, the material accumulated at the dead zone inner edge diffuses more efficiently towards the inner disk, slightly lowering the dust-to-gas ratio in the dust trap. In the high resolution case, the accumulated dust mass between \SI{0.51}{}-\SI{0.6}{au} is \SI{75}{M_\oplus} (in contrast with the \SI{110}{M_\oplus} accumulated in the low resolution case).\\
Other than the slight redistribution of dust, the high resolution simulation behaves in the same way as the standard case, and our conclusions are maintained.\\
We tested that the dust distribution obtained during the dust concentration phase (Phase 1) converges for resolutions resolutions higher than $n_r = 1000$ (tested up to $n_r=3000$), and therefore we do not expect any further changes in the subsequent results after the dead zone reactivation (other than those already reported in this section).

\newpage

\bibliography{main_apj_mgarate.bbl}

\end{document}